\newif\iffigs\figsfalse              
\figstrue                            
\newif\ifbbB\bbBfalse                
\bbBtrue                             

\input harvmac
\def\newdate{11/04/2004}

\newcount\figno \figno=0
\iffigs
 \message{If you do not have epsf.tex to include figures,}
 \message{change the option at the top of the tex file.}
 \input epsf
 \def\fig#1#2#3{\par\begingroup\parindent=0pt\leftskip=1cm\rightskip=1cm
  \parindent=0pt \baselineskip=11pt \global\advance\figno by 1 \midinsert
  \epsfxsize=#3 \centerline{\epsfbox{#2}} \vskip 12pt 
  {\bf Fig. \the\figno:} #1\par \endinsert\endgroup\par }
\else
 \def\fig#1#2#3{\global\advance\figno by 1 \vskip .25in
  \centerline{\bf Figure \the\figno} \vskip .25in}
\fi


\lref\HoriZJ{
K.~Hori, H.~Ooguri and C.~Vafa,
Nucl.\ Phys.\ B {\bf 504}, 147 (1997)
[arXiv:hep-th/9705220].
}

\lref\DouglasSW{
M.~R.~Douglas and G.~W.~Moore,
arXiv:hep-th/9603167.
}

\lref\JohnsonPY{
C.~V.~Johnson and R.~C.~Myers,
Phys.\ Rev.\ D {\bf 55}, 6382 (1997)
[arXiv:hep-th/9610140].
}

\lref\KronheimerZS{
P.~B.~Kronheimer,
J.\ Diff.\ Geom.\  {\bf 29}, 665 (1989).
}

\lref\Mckay{
J.~McKay,
Proc. \ Symp. \ Pure Math. \ {\bf 37}, 183 (1980).}
\lref\KatzFH{
S.~Katz, A.~Klemm and C.~Vafa,
Nucl.\ Phys.\ B {\bf 497}, 173 (1997)
[arXiv:hep-th/9609239].
}

\lref\AspinwallMN{
P.~S.~Aspinwall,
arXiv:hep-th/9611137.
}

\lref\BershadskySP{
M.~Bershadsky, C.~Vafa and V.~Sadov,
Nucl.\ Phys.\ B {\bf 463}, 398 (1996)
[arXiv:hep-th/9510225].
}

\lref\DiaconescuBR{
D.~E.~Diaconescu, M.~R.~Douglas and J.~Gomis,
JHEP {\bf 9802}, 013 (1998)
[arXiv:hep-th/9712230].
}

\lref\KatzEQ{
S.~Katz, P.~Mayr and C.~Vafa,
Adv.\ Theor.\ Math.\ Phys.\  {\bf 1}, 53 (1998)
[arXiv:hep-th/9706110].
}

\lref\CachazoGH{
F.~Cachazo, S.~Katz and C.~Vafa,
arXiv:hep-th/0108120.
}

\lref\CachazoSG{
F.~Cachazo, B.~Fiol, K.~A.~Intriligator, S.~Katz and C.~Vafa,
Nucl.\ Phys.\ B {\bf 628}, 3 (2002)
[arXiv:hep-th/0110028].
}

\lref\KarchYV{
A.~Karch, D.~Lust and D.~J.~Smith,
Nucl.\ Phys.\ B {\bf 533}, 348 (1998)
[arXiv:hep-th/9803232].
}

\lref\HitchinEA{
N.~J.~Hitchin, A.~Karlhede, U.~Lindstrom and M.~Rocek,
Commun.\ Math.\ Phys.\  {\bf 108}, 535 (1987).
}

\lref\GabrielUD{
P.~Gabriel,
Manuscripta Mathematica. \ {\bf 6}, 71 (1972).}

\lref\KacIR{
V.~G.~Kac,
Inv. \ Math. \ {\bf 56}, 263 (1980).}

\lref\KatzGT{
S.~Katz and D.~Morrison,
J. \ Algebraic Geometry {\bf 1}, 449 (1992).}

\lref\OoguriIH{
H.~Ooguri and C.~Vafa,
Nucl.\ Phys.\ B {\bf 500}, 62 (1997)
[arXiv:hep-th/9702180].
}

\lref\ElitzurHC{
S.~Elitzur, A.~Giveon, D.~Kutasov, E.~Rabinovici and A.~Schwimmer,
Nucl.\ Phys.\ B {\bf 505}, 202 (1997)
[arXiv:hep-th/9704104].
}

\lref\BershadskyGX{
M.~Bershadsky, A.~Johansen, T.~Pantev, V.~Sadov and C.~Vafa,
Nucl.\ Phys.\ B {\bf 505}, 153 (1997)
[arXiv:hep-th/9612052].
}

\lref\LindstromPZ{
U.~Lindstrom, M.~Rocek and R.~von Unge,
JHEP {\bf 0001}, 022 (2000)
[arXiv:hep-th/9908082].
}

\lref\DouglasZJ{
M.~R.~Douglas and B.~R.~Greene,
Adv.\ Theor.\ Math.\ Phys.\  {\bf 1}, 184 (1998)
[arXiv:hep-th/9707214].
}

\lref\CalabiAE{E.~Calabi,
Ann. \ Sci. \ Ec. \ Norm. \ Sup. {\bf 12}, 269 (1979).}

\lref\HoriIQ{
K.~Hori and Y.~Oz,
Nucl.\ Phys.\ B {\bf 501}, 97 (1997)
[arXiv:hep-th/9702173].
}

\lref\GreenDD{
M.~B.~Green, J.~A.~Harvey and G.~W.~Moore,
Class.\ Quant.\ Grav.\  {\bf 14}, 47 (1997)
[arXiv:hep-th/9605033].
}

\lref\HananyIE{
A.~Hanany and E.~Witten,
Nucl.\ Phys.\ B {\bf 492}, 152 (1997)
[arXiv:hep-th/9611230].
}

\lref\WittenSC{
E.~Witten,
Nucl.\ Phys.\ B {\bf 500}, 3 (1997)
[arXiv:hep-th/9703166].
}

\lref\OoguriWJ{
H.~Ooguri and C.~Vafa,
Nucl.\ Phys.\ B {\bf 463}, 55 (1996)
[arXiv:hep-th/9511164].
}

\lref\AntoniadisRA{
I.~Antoniadis and B.~Pioline,
Int.\ J.\ Mod.\ Phys.\ A {\bf 12}, 4907 (1997)
[arXiv:hep-th/9607058].
}

\lref\PolchinskiMT{
J.~Polchinski,
Phys.\ Rev.\ Lett.\  {\bf 75}, 4724 (1995)
[arXiv:hep-th/9510017].
}

\lref\WittenEX{
E.~Witten,
Nucl.\ Phys.\ B {\bf 443}, 85 (1995)
[arXiv:hep-th/9503124].
}

\lref\WittenZH{
E.~Witten,
arXiv:hep-th/9507121.
}

\lref\KatzTH{
S.~Katz and C.~Vafa,
Nucl.\ Phys.\ B {\bf 497}, 196 (1997)
[arXiv:hep-th/9611090].
}

\lref\VafaXN{
C.~Vafa,
Nucl.\ Phys.\ B {\bf 469}, 403 (1996)
[arXiv:hep-th/9602022].
}

\lref\SeibergPQ{
N.~Seiberg,
Nucl.\ Phys.\ B {\bf 435}, 129 (1995)
[arXiv:hep-th/9411149].
}

\lref\FiolWX{
B.~Fiol and M.~Marino,
JHEP {\bf 0007}, 031 (2000)
[arXiv:hep-th/0006189].
}

\lref\DouglasQW{
M.~R.~Douglas, B.~Fiol and C.~Romelsberger,
arXiv:hep-th/0003263.
}

\lref\MukaiMF{
S.~Mukai,
Nagoya Math. \ J. {\bf 81}, 153 (1981).}

\lref\HananySJ{
A.~Hanany and A.~Zaffaroni,
JHEP {\bf 9907}, 009 (1999)
[arXiv:hep-th/9903242].
}

\lref\LindstromRT{
U.~Lindstrom and M.~Rocek,
Nucl.\ Phys.\ B {\bf 222}, 285 (1983).
}

\lref\RoblesLlanaDD{
D.~Robles-Llana and M.~Rocek,
arXiv:hep-th/0405230.
}

\lref\AlvarezGaumeVS{
L.~Alvarez-Gaume and D.~Z.~Freedman,
Phys.\ Lett.\ B {\bf 94}, 171 (1980).
}

\lref\FiolAH{
B.~Fiol,
JHEP {\bf 0207}, 058 (2002)
[arXiv:hep-th/0205155].
}

\lref\SeibergPQ{
N.~Seiberg,
Nucl.\ Phys.\ B {\bf 435}, 129 (1995)
[arXiv:hep-th/9411149].
}

\lref\BeasleyZP{
B.~Feng, A.~Hanany and Y.~H.~He,
Nucl.\ Phys.\ B {\bf 595}, 165 (2001)
[arXiv:hep-th/0003085].

C.~E.~Beasley and M.~R.~Plesser,
JHEP {\bf 0112}, 001 (2001)
[arXiv:hep-th/0109053].
}
\lref\ArgyresEH{
P.~C.~Argyres, M.~R.~Plesser and N.~Seiberg,
Nucl.\ Phys.\ B {\bf 471}, 159 (1996)
[arXiv:hep-th/9603042].
}

\lref\BerensteinFI{
D.~Berenstein and M.~R.~Douglas,
arXiv:hep-th/0207027.
}

\lref\EguchiXP{
T.~Eguchi and A.~J.~Hanson,
Phys.\ Lett.\ B {\bf 74}, 249 (1978).
}

\lref\OhBF{
K.~h.~Oh and R.~Tatar,
Adv.\ Theor.\ Math.\ Phys.\  {\bf 6}, 141 (2003)
[arXiv:hep-th/0112040].
}
\lref\OoguriWJ{
H.~Ooguri and C.~Vafa,
Nucl.\ Phys.\ B {\bf 463}, 55 (1996)
[arXiv:hep-th/9511164].
}

\lref\UrangaVF{
A.~M.~Uranga,
JHEP {\bf 9901}, 022 (1999)
[arXiv:hep-th/9811004].
}

\lref\OoguriIH{
H.~Ooguri and C.~Vafa,
Nucl.\ Phys.\ B {\bf 500}, 62 (1997)
[arXiv:hep-th/9702180].
}
\lref\ElitzurFH{
S.~Elitzur, A.~Giveon and D.~Kutasov,
Phys.\ Lett.\ B {\bf 400}, 269 (1997)
[arXiv:hep-th/9702014].
}
\lref\FengMI{
B.~Feng, A.~Hanany and Y.~H.~He,
Nucl.\ Phys.\ B {\bf 595}, 165 (2001)
[arXiv:hep-th/0003085];~

C.~E.~Beasley and M.~R.~Plesser,
JHEP {\bf 0112}, 001 (2001)
[arXiv:hep-th/0109053];~
B.~Feng, A.~Hanany, Y.~H.~He and A.~M.~Uranga,
JHEP {\bf 0112}, 035 (2001)
[arXiv:hep-th/0109063];~
B.~Feng, S.~Franco, A.~Hanany and Y.~H.~He,
JHEP {\bf 0212}, 076 (2002)
[arXiv:hep-th/0205144].
}
 
\lref\WijnholtQZ{
M.~Wijnholt,
Adv.\ Theor.\ Math.\ Phys.\  {\bf 7}, 1117 (2004)
[arXiv:hep-th/0212021].
}

\lref\HerzogQW{
C.~P.~Herzog,
JHEP {\bf 0408}, 064 (2004)
[arXiv:hep-th/0405118].
}
\lref\FrancoMU{
S.~Franco and A.~Hanany,
Fortsch.\ Phys.\  {\bf 51}, 738 (2003)
[arXiv:hep-th/0212299].
}
\lref\FrancoAE{
S.~Franco and A.~Hanany,
JHEP {\bf 0304}, 043 (2003)
[arXiv:hep-th/0207006].
}

\Title{\vbox{\baselineskip12pt
\hbox{YITP-SB-04-58}
\hbox{\tt hep-th/0411059}
}}
{\vbox{\centerline{On ${\cal N}=2$ Seiberg Duality for Quiver Theories}
}}
\smallskip
\medskip\centerline{Daniel Robles-Llana\foot{daniel@insti.physics.sunysb.edu}}
\medskip
\centerline{{\it C.N. Yang Institute for Theoretical Physics,} }
\centerline{{\it State University of New York at Stony Brook,}}
\centerline{{\it NY 11794-3840,}}
\centerline{{\it USA}}

\bigskip
\bigskip

\noindent
We consider a general class of symmetries of hyper-K\"ahler quotients which can be interpreted as classical analogs of Seiberg duality for ${\cal N}=2$ supersymmetric quiver gauge theories in the baryonic Higgs branch. Along the way we find that a limit application of this duality yields new exotic realizations of $ALE$ spaces as hyper-K\"ahler quotients. We also discuss how these results admit in some particular cases a natural interpretation in string theory. We finally comment on a possible relationship with Fourier-Mukai tranforms. 
\Date{\newdate}

\baselineskip 16pt
                  
\newsec{Introduction}
Seiberg duality \SeibergPQ\ is an infrarred equivalence between $N=1$ supersymmetric gauge theories in the conformal window. The original statement is that pure $N=1$ $SU(N_c)$ SQCD with $N_f$ flavors and no superpotential flows to the same infrarred fixed point as $N=1$ supersymmetric gauge theory with gauge group  $SU(N_f-N_c)$, $N_f$ flavors and superpotential $qM\tilde q$, where the mesons $M$ are gauge singlets and can be seen as composites of the original quarks. Since the original formulation of the duality, subsequent work on gauge theories engineered from string theory has made it clear that Seiberg duality can be recast in a larger context. The pioneering works \ElitzurFH\OoguriIH\ showed that the duality could be seen at the level of certain recombinations of branes. These opened the path to more recent works which have systematically exploited the fact that Seiberg dual gauge theories can be engineered by placing $D$ branes on different realizations of the same underlying geometric singularity. In the context of geometric engineering of $ADE$ gauge theories, the dual realizations are related by Weyl reflections of the $ADE$ algebra underlying the geometry \CachazoSG\FiolAH. In the case when the gauge theories can be engineered by placing $D$ branes at singularities admitting a toric description, Seiberg duality was seen to arise as a manifestation of an algebro-geometric symmetry given the name of toric duality \FengMI. Closely related conclusions were also obtained using $(p,q)$ web brane constructions \FrancoAE. More abstract approaches have also been put forward in \BerensteinFI\WijnholtQZ\HerzogQW.

The lesson to be drawn from these works is that there is a close connection between algebraic-geometric quotients and field theory dualities. In the case of $N=1$ supersymmetry the relevant one is a K\"ahler quotient, and indeed toric duality can be seen as a symmetry relation two different K\"ahler quotients performed in the toric setup. This suggests the possibility of an $N=2$ Seiberg duality, which should in this case arise through some symmetry of hyper-K\"ahler quotients. This is the main point of the present paper. In the case of $SU(N_c)$ gauge group and $N_f$ flavors this was already oberved in \ArgyresEH\ and \AntoniadisRA. At the level of $N=2$ quiver theories with $ADE$ gauge group this was implicit in the treatment of \CachazoSG. In general, the main geometric property that one can exploit for four dimensinal $N=2$ theories is the fact that the Higgs branch is a hyper-K\"ahler manifold, and is not corrected by quantum corrections. This enables one to construct the Higgs branch of arbitrary $N=2$ quiver theories through the hyper-K\"ahler quotient procedure \HitchinEA. In this approach one can verify directly that the hyper-K\"ahler potentials for the baryonic Higgs branches of two $N=2$ Seiberg dual theories match exactly \RoblesLlanaDD. In addition, the mapping between the geometric parameters (FI parameters in the gauge theory)  reproduces exactly what one would obtain through other approaches, giving it a clean and direct interpretation. Also, and in contrast to other approaches involving $D$ branes at singularities, the method completely bypasses the issue of engineering the gauge theory from the geometry, a process which in some cases can be very involved. Moreover, we believe that this approach to the $N=2$ duality captures some essential aspects and suggests an extension to the $N=1$ case, which could in priciple be generalized to any singularity, independently of string theory.

In this work, following \RoblesLlanaDD, we systematically exploit the hyper-K\"ahler quotient procedure to analyze the Higgs branch of arbitrary $N=2$ quiver theories. We follow a constructive approach, starting in Section 2 with a unified treatment of $ADE$ singularities, $ALE$ spaces and their hyper-K\"ahler quotient construction, in such a way that the generalization to arbitrary non-chiral quiver moduli spaces is straightforward. In Section 3 we discuss $N=2$ Seiberg duality as a symmetry of hyper-K\"ahler quotients obtained as moduli spaces of non-chiral quivers. Along the way we discuss some exotic examples from the field theoretical point of view, but which appear naturally from the algebraic-geometric standpoint. In particular we analyze the case in which $N=2$ Seiberg duality yields a null rank for the dual gauge group, which in some particular application allows us to generalize Kronheimer's construction of $ALE$ spaces. In Section 4 we discuss geometric engineering and generalized Hanany-Witten setups in order to, in Section 5, see the implications of our general discussion in the case of gauge theories constructed from string theory. We end this work, in Section 6, with some curious connection between $N=2$ Seiberg duality and Fourier-Mukai transforms.

\newsec{$ADE$ singularities, $ALE$ spaces and hyper-K\"ahler quotients}

\subsec{ADE singularities and their resolutions}
Quotients of ${\bf C}^2$ by a discrete group $\Gamma$ of $SU(2)$ can be described by equations $f(x,y,z)=0$ in ${\bf C}^3$ and admit an $ADE$ classification. Depending on whether $\Gamma$ is cyclic ($A$ cases), dyhedral ($D$ cases) or tetrahedral, octahedral and dodecahedral ($E$ cases) these are
\eqn\twentytwo{\eqalign{
& A_r ~:~~~~xy+z^{r+1}=0\,, \cr 
& D_r ~:~~~~x^2 +y^2 z + z^{r-1}=0\,, \cr
& E_6 ~:~~~~x^2 + y^3 + z^4 = 0\,, \cr 
& E_7 ~:~~~~x^2 + y^3 + yz^3 = 0\,, \cr 
& E_8 ~:~~~~x^2 + y^3 + z^5 = 0\,.
}}
These are singular at $x=y=z=0$, but can be made smooth by deforming the above polynomials to $f(x,y,z;t_i)$ in such a way that there is no point in ${\bf C}^3$ satisfying $f=|df|=0$ (see, {\it e.g.} \HoriZJ\CachazoGH\KatzGT). There exist $r$ such deformations for $ADE$ group of rank $r$. For the $A$ and $D$ series the deformed equations are
\eqn\tthree{\eqalign{
& A_r ~:~~x^2+y^2+\prod_{i=1}^{r+1} (z+t_i) = 0,\cr 
& D_r ~:~~x^2+y^2 z + {\prod^{r}_{i=1} (z+ t^{2}_{i}) - \prod^{r}_{i=1} t^{2}_{i} \over z} + 2 \prod^{r}_{i=1} t_{i} y = 0\,,}}
and more complicated expressions for the $E$ cases.

The deformation process can be achieved by a series of sequential blow-ups. In this way one builds a map $\tilde S\rightarrow S$ between the smooth resolution $\tilde S$ and the singular space $S$. This map is an isomorphism away from the singular point. The inverse image of the singular point is, however, given by a set of ${\bf CP^1}$'s which intersect according to the adjacency matrix of the corresponding $ADE$ group \Mckay. In other words, for each of the $ADE$ singularities, to each ${\bf CP}^1$ of the blow-up we assign a simple root of the simply laced Lie group. Its Dynkin diagram then just tells us how these spheres intersect. Moreover, the holomorphic volumes $\zeta_i$ of these spheres are related to the deformation parameters as
\eqn\twentyfive{\eqalign{
& A_r ~:~~~\zeta_i = t_i - t_{i+1}\,,~~~~i=1,\ldots,r~,\cr 
& D_r ~:~~~\zeta_i = t_i - t_{i+1}\,,~~~~i=1\ldots,r-1~,~~~\zeta_{r}=t_{r-1} + t_r ~.
}}
When the $ADE$ group is promoted to its affine extension (that means including the simple root of the trivial representation), the resulting smooth resolution is diffeomorphic to an $ALE$ space. These are a class of four dimensional hyper-K\"ahler manifolds which at infinity approach ${\bf R}^4/\Gamma$, where $\Gamma$ is the expected discrete subgroup of $SU(2)$. These were systematically constructed by Kronheimer \KronheimerZS\ through the hyper-K\"ahler quotient procedure \HitchinEA, to which we turn next.

\subsec{The hyper-K\"ahler quotient}
The hyper-K\"ahler quotient construction is a powerful technique to generate new hyper-K\"ahler manifolds from known ones. We remind the reader that a $4n$ real dimensional Riemannian manifold is hyper-K\"ahler if it is endowed with three covariantly constant complex structures (with respect to the Levi-Civita connection) $I,J,K$ satisfying the quaternionic algebra
\eqn\defhyp{
I^2=J^2=K^2=-1,~~~~IJ=-JI=K~.}
Equivalently, it has holonomy contained in $Sp(n)$. 

Suppose now one is given a hyper-K\"ahler manifold $M$ with a group of isometries $G$ acting freely on $M$ and preserving the three complex structures. \foot{These are called {\it triholomorphic}.} When contracted with the K\"ahler forms derived from the three complex structures (that is, $\omega_i(X,Y)= g(I_iX,Y)$ for $I_1=I,I_2=J,I_3=K$) vector fields $X$ generating ${\cal G}$, the Lie algebra of G, give rise to $3\cdot {\rm dim}(G)$ {\it moment maps} as
\eqn\defmomentmaps{\eqalign{
0={\cal L}_X\omega_i=i(X)d\omega_i+d(i(X)\omega_i) ~~~~&\Rightarrow d(i(X)\omega_i)=0\cr
& \Rightarrow i(X)\omega_i = d\mu^X_i~.}}
The moment maps $\mu^X_i$ take points in $M$ to ${\cal G}^*\otimes {\bf R}^3$, where ${\cal G}^*$ is the dual Lie algebra of G. The hyper-K\"ahler quotient of $M$ by $G$ is then
\eqn\hykqdef{
{\cal M}=\left[ \mu^{-1}_1(\zeta_1)\cap\mu^{-1}_2(\zeta_2)\cap\mu^{-1}_3(\zeta_3)\right]/G}
where $\zeta_i$ are arbitrary central elements in $G$ (this ensures that $G$ acts on the subspaces $\mu^{-1}_i(\zeta_i)$), and are called the levels of the hyper-K\"ahler quotient. One of the main results in \HitchinEA\ is the proof that ${\cal M}$ is itself a hyper-K\"ahler manifold of dimension $4n-4\cdot{\rm dim}(G)$.

To finish this subsection, let us mention a form of \hykqdef\ which is more familiar from the physics literature. On $M$ one can pick a preferred complex structure, $I$ say, and write \hykqdef\ in a coordinate system which makes this complex structure manifest. The new moment maps can then be written in complex notation as
\eqn\nmm{
\mu_+=\mu_2 + i \mu_3~,~~~~\mu_{R}=\mu_1~,~~~~\mu_-=\mu_2-i\mu_3=\bar \mu_+~,}
and
\eqn\nnm{
{\cal M}=\left[ \mu_+(\zeta_+)\cap\mu_{R}(\zeta_R)\cap\mu_-(\zeta_-)\right]/G~,}
where
$\zeta_+=\zeta_2 + i\zeta_3$, $ \zeta_R = \zeta_1$, and $\zeta_-=\bar\zeta_+$.

\subsec{Kronheimer's construction}
In \KronheimerZS\ Kronheimer was able to prove that $ALE$ spaces arise as a particular instance of the hyper-K\"ahler quotient construction described above. We summarize here his construction. Take $\Gamma$ to be a finite subgroup of $SU(2)$, and let $Q$ be the defining two-dimensional representation acting on ${\bf C}^2$. Moreover let $R_0, R_1,\ldots,R_r$ be irreducible representations of $\Gamma$, with $R_0$ the trivial representation. The product $Q\otimes R_i$ can be decomposed into irreducibles as
\eqn\irrdec{
Q\otimes R_i = \oplus a_{ij}R_j~.}
The McKay correspondence \Mckay\ asserts that to any given $\Gamma\in SU(2)$ one can associate a simply laced Lie group $G$ in such a way that the Clebsch-Gordan coefficients $a_{ij}$ are the entries of the adjacency matrix of its {\it extended} Dynkin diagram, where the extended node corresponds to the negative of the highest root
\eqn\higest{
e_0=-\sum_{i=1}^re_i~.}
With this in mind, the statement of Kronheimer is that the $ALE$ space which aymptotes to ${\bf R}^4/\Gamma$ can be constructed as the hyper-K\"ahler quotient of flat \foot{And thus trivially hyper-K\"ahler if the real dimension is a multiple of four.} space by a suitably chosen action of $G$. This parent hyper-K\"ahler space is
\eqn\parenthyp{\eqalign{
M&={\rm Hom}_\Gamma(R,Q\otimes R)\cr
&=\bigoplus_{i,j}\left[ {\rm Hom}_\Gamma\left(R_i,Q\otimes R_j\right)\otimes{\rm Hom}\left({\bf C}^{N_i},{\bf C}^{N_j}\right)\right]\cr
&=\bigoplus_{i,j}a_{ij}{\rm Hom}({\bf C}^{N_i},{\bf C}^{N_j})~,}}
where $N_i={\rm  dim}(R_i)$ and we have made use of  \irrdec\ and Schur's lemma. The group one quotients by is $G=\bigotimes_i U(N_i)$. Denoting an element of ${\rm Hom}({\bf C}^{N_i},{\bf C}^{N_j})$ by $\Phi_{ij}$, the holomorphic and K\"ahler forms on $M$ can be written as 
\eqn\holkah{\eqalign{
\Omega_+&=\sum_{i>j}a_{ij}{\rm Tr}\left[d\Phi_{ij}\wedge d\Phi_{ji}\right]~,\cr
\omega&~=\sum_{i\neq j}a_{ij}{\rm Tr}\left[d\Phi_{ij}\wedge d\bar\Phi_{ij}\right]~,\cr
\Omega_-&=\bar\Omega_+~,}}
where, as in the last paragraph of the last subsection we have chosen a preferred complex structure for which $\Phi_{ij}$ is holomorphic and $\bar\Phi_{ij}$ antiholomorphic. The group $G$ acts on $\Phi_{ij}$ in the natural way: $\Phi_{ij}$ tranforms in the $(1,\ldots,\bar N_i,\ldots,N_j,\ldots,1)$ of $G$. One then has moment maps generated by this action  and $ALE$ spaces are obtained as the hyper-K\"ahler quotients
\eqn\alequot{
{\cal M}_\zeta=\left[\mu^{-1}_+(\zeta_+)\cap\mu^{-1}_R(\zeta_R)\cap \mu^{-1}_-(\zeta_-)\right]/G~.}
Here $\zeta_\pm=(\zeta_{\pm 0},\zeta_{\pm 1},\ldots,\zeta_{\pm r})$ (and similarly for $\zeta_R$), are central elements of $G$ and can be seen as coming from the $U(1)$ factors in $\bigotimes_iU(N_i)$. They are moduli of the $ALE$ spaces, and correspond to the sizes of the ${\bf CP}^1$'s of the blown-up affine geometry.\foot{However, by taking traces of the moment maps they can be seen to obey certain constraints.}

\subsec{Superspace description and generalization}
The construction of the previous subsection has a very natural interpretation in physical terms, as it has been known for a long time that $N=2$ supersymmetric sigma-models are an ideal framework for performing hyper-K\"ahler quotients \AlvarezGaumeVS\LindstromRT\HitchinEA. Moreover the extension to arbitrary non-chiral quiver theories is immediate. 

Consider an arbitrary quiver with $r$ nodes and all links bidirectional. As in the case of simply-laced Lie groups we have a linear $r$ dimensional ``root'' space for which we can choose a basis $\{e_1,e_2,\ldots,e_r\}$. Any quiver diagram of this sort can be labeled by a vector $\vec v=N_1e_1+N_2e_2+\ldots+N_re_r$, $N_i\geq 0$ and a symmetric matrix of integers $[A]=a_{ij}$ specifying the number of links between nodes $i$ and $j$. We can then consider the hyper-K\"ahler quotient of
\eqn\generalquiver{
M=\bigoplus_{i,j}a_{ij}{\rm Hom}({\bf C}^{N_i},{\bf C}^{N_j})}
by $G=\bigotimes_i U(N_i)$. The complex dimension of the quotient for a quiver labeled by vector $\vec v$ and adjacency matrix $a_{ij}$ can be shown to be $D=2-v^t C v$, where $[C]_{ij}=2\delta_{ij}-a_{ij}$ is the associated Cartan matrix \KacIR.

In physical terms, the space $M$ is spanned by hypermultiplet superfields $(\Phi_{ij}^{m},\Phi_{ji}^{m})$, where $m$ runs from 1 to $a_{ij}$\foot{Of course, we understand there are no such hypermultiplet fields if $a_{ij}=0$, that is if nodes $i$ and $j$ are not connected.}. The group $G=\bigotimes_i U(N_i)$ is the gauge group. In $N=1$ language $\Phi_{ij}^{m}$ and $\Phi_{ji}^m$ are chiral superfields transforming in the $(\bar N_i, N_j)$ and $(\bar N_j, N_i)$ representations of $U(N_i)\otimes U(N_j)$ respectively. Written in terms of these superfields the $N=2$ supersymmetric superspace lagrangian is
\eqn\superspaceaction{\eqalign{
{\cal L}&=\int d^4 \theta ~ \left[ \sum_{m=1}^{a_{ij}}\sum_{i\neq j} {\rm Tr} \left( \Phi_{ij}^{m}e^{-V_j}\bar\Phi_{ij}^{m}e^{V_i}\right) ~-~\sum_{i}c_i{\rm Tr}~V_i\right] \cr
&+\int d^2\theta \left[ \sum_{m=1}^{a_{ij}}\sum_{i,j} {\rm Tr} \left(s_{ij}\Phi_{ji}^m S_i\Phi_{ij}^m \right) ~-~\sum_i b_i{\rm Tr}~S_i\right]~+({\rm h.c.})}}
where $s_{ij}$ is defined as $\pm 1$ if nodes $i$ and $j$ are linked, and $s_{ij}=-s_{ji}$. The $S_i$'s are auxiliary fields. Their equations of motion give the holomorphic moment maps
\eqn\complexmomentmaps{
\sum_{m=1}^{a_{ij}}\sum_{j}s_{ij}\Phi_{ij}^{m}\Phi_{ji}^{m}=b_i~.}
The $V_i$'s are gauge superfields, and their equations of motion give the real gauged moment maps
\eqn\gaugedrealmomentmaps{
\sum_{m=1}^{a_{ij}}\sum_{j}\left(\bar\Phi_{ij}^{m}e^{-V_j}\Phi_{ij}^{m}e^{V_i} -e^{-V_i}\Phi_{ij}^{m}e^{V_j}\bar\Phi_{ij}^{m}\right)=c_i~,}
where $b_i \equiv\zeta_{+i}$, $c_i\equiv\zeta_{Ri}$ are $FI$ parameters. Solving \gaugedrealmomentmaps\ for $V_i$, subject to \complexmomentmaps, and fixing the {\it complex} gauge invariance yields the horizontal hyper-K\"ahler quotient metric when substituting back in \gaugedrealmomentmaps. This is the superspace version of an ordinary quotient. In the normal case one couples the scalar fields parameterizing the coordinates of the target space to a gauge connection enforcing the gauge invariance by covariantizing the derivatives. Then one restricts oneself to a gauge slice by solving the equations of motion for the connection and substituting back into the action. The gauge coupling in superspace is given by the first line in \superspaceaction. The main feature of superspace is that the gauge invariance is naturally complexified, as the ordinary gauge parameters are lowest components of chiral superfields. In the case of $N=1$ supersymmetry this means that one can perform a K\"ahler quotient as an ordinary symplectic quotient but with respect to a {\it complexified} gauge invariance. \foot{And restrict oneself to the set of {\it stable} orbits.} In other words one does not need to solve the real moment map equations. In the $N=2$ case, the hyper-K\"ahler quotient, one needs only solve the F-term equations \complexmomentmaps\ and again divide by the {\it compexified} gauge group \HitchinEA.

\newsec{Quotient symmetries: Weyl reflections and Seiberg duality}
\subsec{Weyl reflections in root space} 
We have seen in the previous section that a non quiral quiver can be characterized by a vector $\vec v$ labeling the indices at the nodes and a symmetric adjacency matrix $a_{ij}$ specifying the number of links connecting nodes $i$ and $j$. The vector $\vec v$ lives in a vector space ${\cal V}$ and one might wonder whether there exist linear transformations in ${\cal V}$ that yield equivalent hyper-K\"ahler quotients. 

In the context of geometrically engineered $N=1$ $ADE$ gauge theories, it was noticed in \CachazoSG\ that Weyl reflections around primitive roots of the $ADE$ group underlying the geometry correspond to a change of basis for the cycles that D-branes are wrapping, and should thus lead to dual gauge theories. The relevant gauge theory duality is Seiberg duality \SeibergPQ. The autors of \CachazoSG\ consider the case of $N=2$ supersymmetric theories deformed to $N=1$ by addition of a polynomial superpotential. The underlying principle of the $N=1$ gauge theory duality is really a symmetry of non chiral quivers, and is not specific to $ADE$ algebras. This was emphasized in \FiolAH.
     
This symmetry can be easily phrased in the root space, as the set of linear transformations of the basis of generalized roots which leaves the quiver moduli space (the hyper-K\"ahler quotient associated to it) invariant (modulo a redefinition of the resolution parameters).  These are given by
\eqn\ltonroots{
e_i\rightarrow e'_i=-e_1,~~~~e_j\rightarrow e'_j=e_j+a_{ji}e_i,~~~~~(j\neq i),}
for any $i$. The coordinates identifying the given quiver in the new basis are found by imposing $\vec v=\sum_{i=1}^{r}N_ie_i=\sum_{i=1}^{r}N'_ie'_i=\vec v'$, which yields
\eqn\ltdegrees{
N_i\rightarrow N'_i=\sum_j a_{ij}N_j -N_i,~~~~N'_j=N_j,~~~~~(j\neq i)~.}
Finally, the resolution parameters ($FI$ parameters) undergo the same transformations as the simple roots, namely
\eqn\restr{
(b_i,c_i,\bar b_i)\rightarrow(b_i,c_i,\bar b_i)- (2\delta_{ij} -a_{ij})(b_j,c_j,\bar b_j)~.}
Interpreting the nodes adjacent to $i$ as flavors charged under $U(N_i)$ \ltdegrees\ gives the same transformation of the gauge group as Seiberg duality $N_c\rightarrow N_f-N_c$.

The transformations \ltonroots,\ltdegrees,\restr\ should be directly verifyable at the level of the hyper-K\"ahler potentials derived through the hyper-K\"ahler procedure. This should prove the metric equivalence of arbitrary Seiberg dual $N=2$ quiver gauge theories on the Higgs branch. We turn to the explicit proof next.

\subsec{$N=2$ Seiberg duality as an equivalence of hyper-K\"ahler quotients}
The problem can be formulated in its most general terms as proving that the hyper-K\"ahler quotient associated to the two quivers in Fig. 1 is the same, modulo a redefinition of the FI parameters. We consider first the case in which all nodes are gauged \RoblesLlanaDD. \foot{The case of $U(N_c)$ gauge theory with $N_f$ massless flavors was already considered in \ArgyresEH\AntoniadisRA.} The proof in this case makes the inlcusion of massless flavors straightforward. \fig{Two $N=2$ Seiberg dual quivers. The duality is performed around node $I$. Node $I$ is connected to node $i_k$ by $a_{Ii_k}$ bidirectional links.}{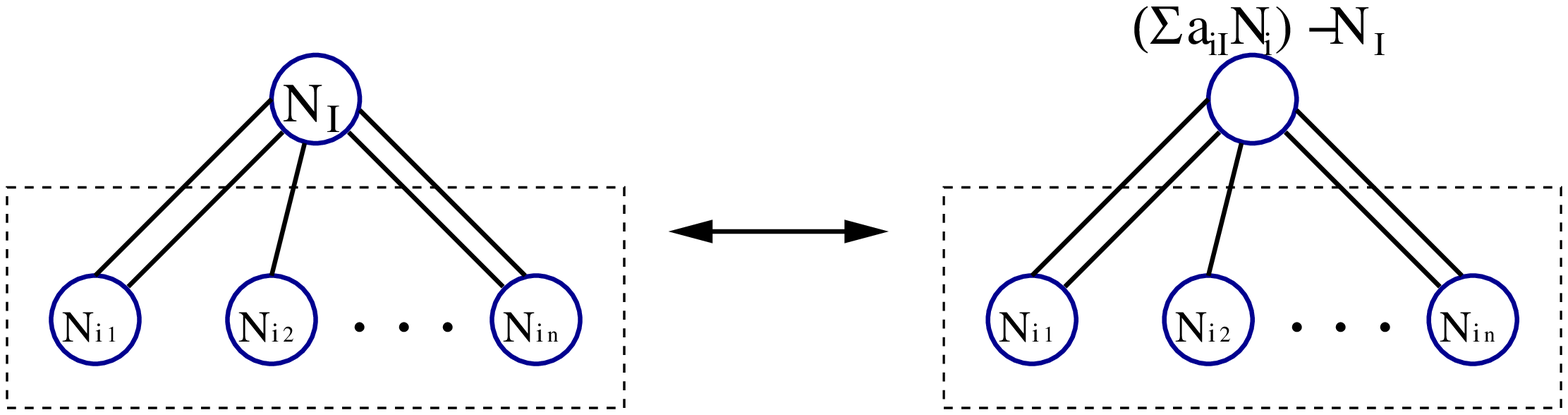}{140mm} Following the discussion in Section 2 the respective superspace lagrangians for the two quivers in Fig. 1 are given by
\eqn\eon{\eqalign{
{\cal L}=& \int d^4\theta~ \left[ \sum_{m=1}^{a_{Ij}}\sum_{j}{\rm Tr}\left( \bar\Phi_{Ij}^m e^{V_I} \Phi_{Ij}^m e^{-V_j}+\bar\Phi_{jI}^m e^{V_j} \Phi_{jI}^m e^{-V_I}    \right)\right]~+\cr
& + \int d^4\theta~\left[-c_I  {\rm Tr} ~V_I-\sum_{i\neq I, a_{Ii}\neq 0}c_i{\rm Tr}~V_i \right] ~+\cr
& + \int d^2\theta~\left[\sum_{m=1}^{a_{ij}}\sum_{j} {\rm Tr} \left(s_{Ij}\Phi_{jI}^m S_I\Phi_{Ij}^m+s_{jI}\Phi_{Ij}^m S_j \Phi_{jI}^m \right) \right]~ +~({\rm h.c.})~+\cr
&+ \int d^2\theta~\left[- b_I{\rm Tr}~ S_I -\sum_{i\neq I,a_{Ii}\neq 0}b_i{\rm Tr}~S_i\right]~+~({\rm h.c.})~+\cr
&+ {\cal L}(extra)~,}}
and
\eqn\eonbis{\eqalign{
\tilde{\cal L}=& \int d^4\theta~ \left[ \sum_{m=1}^{a_{Ij}}\sum_{j}{\rm Tr}\left( \bar{\tilde\Phi}_{Ij}^m e^{V_I} \tilde\Phi_{Ij}^m e^{-V_j}+\bar{\tilde\Phi}_{jI}^m e^{V_j} \tilde\Phi_{jI}^m e^{-V_I}    \right)\right]~+\cr
& + \int d^4\theta~\left[-\tilde c_I  {\rm Tr} ~\tilde V_I-\sum_{i\neq I, a_{Ii}\neq 0}\tilde c_i{\rm Tr}~V_i \right] ~+\cr
& + \int d^2\theta~\left[\sum_{m=1}^{a_{ij}}\sum_{j} {\rm Tr} \left(s_{Ij}\tilde\Phi_{jI}^m \tilde S_I\tilde \Phi_{Ij}^m+s_{jI}\tilde\Phi_{Ij}^m S_j \tilde \Phi_{jI}^m \right) \right]~ +~({\rm h.c.})~+\cr
&+ \int d^2\theta~\left[- \tilde b_I{\rm Tr}~ \tilde S_I -\sum_{i\neq I,a_{Ii}\neq 0}\tilde b_i{\rm Tr}~S_i\right]~+~({\rm h.c.})~+\cr
&+ {\cal L}(extra)~.}}
where $V_I$ and $\tilde V_I$ are $U(N_I)$ and $U(\tilde N_I)$ gauge fields, $\tilde N_I=\sum a_{Ii}N_i -N_I$ and
\eqn\lextra{\eqalign{
{\cal L}(extra) =\int d^4\theta& \left[ \sum_{m=1}^{a_{ij}}\sum_{i,j\neq I} {\rm Tr} \left( \bar\Phi_{ij}^m e^{V_i} \Phi_{ij}^m e^{-V_j} \right)  - \sum_{i\neq I,a_{Ii}=0} c_i {\rm Tr}~ V_i \right] +\cr
& + \int d^2\theta~\left[\sum_{m=1}^{a_{ij}}\sum_{i,j\neq I}{\rm Tr} \left(\Phi_{ji}^m S_{i}\Phi_{ij}^m \right)-\sum_{i\neq I,a_{Ii}=0} b_i {\rm Tr} ~S_i\right]+({\rm h.c.}) ~.}}
is the part coming from the nodes not connected to $I$ and plays a spectator role in what follows. The discussion can be further simplified if one manages to reduce it to the case in which $I$ is connected to a single node. For this define the $M\times M$ block-diagonal matrix (where $M=(\sum_j a_{Ij}N_j)$)
\eqn\jive{
V = \left( \matrix{ V_{i_1}^{\otimes a_{Ii_1}} & \ldots & 0 \cr \vdots & \ddots & \vdots \cr 0 & \ldots & V_{i_n}^{\otimes a_{Ii_n}}  } \right)~, }
where $V_{i_k}$ appears $a_{Ii_k}$ times. Further define the $N_I \times M$ and $M\times N_I$ matrices
\eqn\ssxix{
\Phi_+ = (\Phi_{Ii_1}^1, \ldots ,\Phi_{li_1}^{a_{Ii_1}},\ldots, \Phi_{Ii_m}^{a_{Ii_m}}) ,~~~~~~ \Phi_- =\left(\matrix{\Phi_{i_1 I}^1\cr ,\vdots \cr\Phi_{i_m I}^{a_{Ii_m}}}\right). }
One can then rewrite the lagrangian \eon\ as
\eqn\sssix{\eqalign{
{\cal L} =  \int & d^4\theta~  \left[ \rm{Tr} \left( \bar\Phi_+ e^{V_I} \Phi_+ e^{-V} +  \bar\Phi_- e^{V} \Phi_- e^{-V_I}\right) - ~ c_I {\rm Tr} ~V_I -\sum_{i\neq I ,a_{Ii}\neq0}c_i{\rm Tr}~V_i\right] + \cr
& + \int d^2\theta~\left[{\rm Tr}~\left(\Phi_- S_I \Phi_+ - \sum_{m=1}^{a_{Ii}}\sum_i \Phi_{Ii}^mS_i\Phi_{iI}^m\right) -~ b_I{\rm Tr}~S_I-\sum_{i\neq I,a_{Ii}\neq 0}b_i{\rm Tr}~S_i\right] \cr & +({\rm h.c.}) 
+{\cal L}(extra)~}}
Similarly, defining the $M\times \tilde N_I$ and $\tilde N_I\times M$ matrices 
\eqn\tildes{
\tilde \Phi_+=\left(\matrix{\tilde\Phi_{i_1 I}^1\cr\vdots\cr\tilde\Phi_{i_m I}^{a_{Ii_n}}}\right),~~~~~~\tilde \Phi_- =(\tilde\Phi_{I i_1}^1,\ldots,\tilde\Phi_{I i_1}^{a_{Ii_1}},\ldots,\tilde\Phi_{I i_m }^{a_{Ii_n}})~,}
an analogous rewriting applies to the dual superspace lagrangian \eonbis, which becomes
\eqn\duall{\eqalign{
\tilde{\cal L}=\int  d^4\theta~&  \left[ {\rm Tr} \left( \bar{\tilde\Phi}_- e^{\tilde V_I} {\tilde\Phi}_- e^{-V} +  \bar{\tilde\Phi}_+ e^{V} {\tilde\Phi}_+ e^{-\tilde V_I}\right) - ~ \tilde c_I {\rm Tr} ~\tilde V_I -\sum_{i\neq I,a_{Ii}\neq 0}~\tilde c_i{\rm Tr}~V_i\right] + \cr
& + \int d^2\theta~\left[{\rm Tr}\left({\tilde\Phi}_+ \tilde S_I {\tilde\Phi}_- -\sum_{m=1}^{a_{Ii}}\sum_i \tilde\Phi_{Ii}^mS_i\tilde\Phi_{iI}^m\right) -~ \tilde b_I{\rm Tr}~\tilde S_I-\sum_{i\neq I,a_{Ii}\neq 0}\tilde b_i{\rm Tr}~ S_i\right] \cr & +({\rm h.c.}) + {\cal L}(extra)~.}}
We are now ready to explicitely perform the part of the two quotients coming from node $I$. 

Varying the kinetic terms in \sssix\ and \duall\ with respect to $V_I$ and $\tilde V_I$ we obtain the D-flatness equations
\eqn\ssss{\eqalign{ 
M_+ ~ e^{V_I}~-~ e^{-V_I}M_- &=c_I {\bf 1}_{N_I}~,\cr
e^{\tilde V_I}~\tilde M_- ~-~\tilde M_+ e^{-\tilde V_I}&=\tilde c_I \tilde{\bf 1}_{\tilde N_I} ~, }}
where $M_+ =\Phi_+ e^{-V}\bar\Phi_+$, $M_-=\bar\Phi_- e^V \Phi_-$, and $\tilde M_-=\tilde\Phi_-e^{-V}\bar{\tilde\Phi}_-$, $\tilde M_+=\bar{\tilde\Phi}_+ e^V \tilde\Phi_+$. Similarly, varying with respect to $S_I$, $\tilde S_I$ yields the F-flatness equations
\eqn\xxxxx{\eqalign{\Phi_+\Phi_- &= b_I~,\cr\tilde\Phi_-\tilde\Phi_+&=\tilde b_I~.
}}
The solution to the D-term equations \ssss\ are
\eqn\sossss{\eqalign{
e^{V_I}={1\over 2}M_+^{-1}\left(c_I+\sqrt{c_I^2+4M_+M_-}\right)~,\cr
e^{-\tilde V_I}={1\over 2}\tilde M_+^{-1}\left(-\tilde c_I+\sqrt{\tilde c_I^2+\tilde M_+\tilde M_-}\right)~.}}
Plugging back into \sssix\ and \duall\ one gets
\eqn\rwe{\eqalign{
{\cal L}=\int d^4\theta&~\left[{\rm Tr}\left(\sqrt{c_I^2+4 M_+ M_-}\right)-c_I{\rm Tr}~{\rm ln}\left(c_I+\sqrt{c_I^2+4M_+M_-}\right)\right]~+\cr
&+\int d^4\theta~\left[ c_I{\rm Tr}~{\rm ln}~M_+-\sum_{i\neq I,a_{Ii}\neq 0}c_i{\rm Tr}~V_i \right]~ + \cr
&+\int d^2\theta~\left[{\rm Tr}\left(\sum_{m=1}^{a_{iI}}\sum_i\Phi_{Ii}S_i\Phi_{iI}\right)-\sum_{i\neq I, a_{Ii}\neq 0}b_i{\rm Tr}~S_i\right]+({\rm h.c.})+\cr
&+{\cal L}(extra)~,}}
and
\eqn\rwet{\eqalign{
\tilde{\cal L}=\int d^4\theta&~\left[{\rm Tr}\left(\sqrt{\tilde c_I^2+4\tilde M_+\tilde M_-}\right)+\tilde c_I{\rm Tr}~{\rm ln}\left(-\tilde c_I+\sqrt{\tilde c_I^2+4\tilde M_+\tilde M_-}\right)\right]~-\cr
&-\int d^4\theta~\left[\tilde c_I{\rm Tr}~{\rm ln}~\tilde M_++\sum_{i\neq I,a_{ii}\neq 0}\tilde c_i{\rm Tr}~V_i\right] + \cr
&+\int d^2\theta~\left[{\rm Tr}\left(\sum_{m=1}^{a_{iI}}\sum_i\tilde\Phi_{Ii}S_i\tilde\Phi_{iI}\right)-\sum_{i\neq I, a_{Ii}\neq 0}\tilde b_i{\rm Tr}~S_i\right]+({\rm h.c.})+\cr
&+{\cal L}(extra)~,}}
where $M_{\pm}$ and $\tilde M_{\pm}$ are subject to \xxxxx. We can use the $U(N_I)$ and $U(\tilde N_I)$ (where $\tilde N_I=\left(\sum_i a_{iI}N_i\right) -N_I$) to bring $\Phi_+$ and $\tilde\Phi_+$ to the form
\eqn\soc{\eqalign{
\Phi_+ &=\left(\matrix{{\bf 1}_{N_I\times N_I} & Q_{N_I\times \tilde N_I}}\right),~~~\tilde\Phi_+=\left(\matrix{P_{N_I\times \tilde N_I} \cr {\bf 1}_{\tilde N_I\times \tilde N_I}}\right)~.\cr}} 
The elements of $Q$ and $P$ are part of the coordinates of the respective quotients. The rest of the coordinates are found by solving the F-term constraints for $\Phi_-$ and $\tilde \Phi_-$. This yields
\eqn\tf{
\Phi_-=\left(\matrix{b_I{\bf 1}-QB \cr B}\right),~~~\tilde\Phi_-=\left(\matrix{C & \tilde b_I{\bf 1} - CP}\right)~,}
where $B$ and $C$ are $\tilde N_I\times N_I$ matrices. With these we can write
\eqn\mmi{
\Phi_-\Phi_+=\left(\matrix{b_I{\bf 1}-QB & b_IQ-QBQ \cr B & BQ}\right)~,~~~~\tilde\Phi_+\tilde\Phi_-=\left(\matrix{PC & \tilde b_IP-PCP \cr C & \tilde b_I{\bf 1}-CP}\right)~.}
The gist of the proof is now whether there exists a coordinate transformation relating $\Phi_{\pm}$ to $\tilde\Phi_{\pm}$ such that the quotient hyper-K\"ahler potentials \rwe\ and \rwet\ are mapped to each other. We will see that this transformation is $Q=-P$, $B=C$. With these choices the meson matrices \tf\ and \mmi\ become 
\eqn\messonmattr{
\Phi_-\Phi_+=\left(\matrix{b_I{\bf 1}-QB & b_IQ-QBQ \cr B & BQ}\right)~,~~~~
\tilde\Phi_+\tilde\Phi_- =\left(\matrix{-QB & -\tilde b_IQ-QBQ \cr B & \tilde b_I{\bf 1}+BQ}\right)~.}
Defining the projectors $P_N=\left(\matrix{{\bf 1}_{N_I\times N_I} & 0 \cr 0 & 0}\right)$, $\tilde P_I=\left(\matrix{0 & 0 \cr 0 & {\bf 1}_{\tilde N_I\times \tilde N_I}}\right)$, we see that provided  $b_I=-\tilde b_I$, we can write
\eqn\mess{
\Phi_+\Phi_-=b_I e^{i\Lambda}P_Ne^{-i\Lambda}~,~~\tilde\Phi_+\tilde\Phi_-=b_Ie^{i\Lambda}P_Me^{-i\Lambda}~,~~e^{i\Lambda}\left(\matrix{{\bf 1}-{1\over b_I}QB & -Q \cr {1\over b_I}B & {\bf 1}}\right).}
The trace of any function of $M_+M_-$  can then be written as
\eqn\trs{
{\rm Tr}f(M_+M_-)~=~{\rm Tr}f\left(|b_I|^2P_N\left(e^{-i\Lambda}e^{-V}e^{i\bar\Lambda}\right)P_N\left(e^{-i\Lambda}e^Ve^{i\Lambda}\right)\right)~.}
with a similar expression for $\tilde M_+\tilde M_-$ with $P$ interchanged with $\tilde P$. Using then the matrix identity (valid for any invertible matrix ${\cal O}$) 
\eqn\lk{
{\rm Tr}f(P_I{\cal O}P_I{\cal O}^{-1})={\rm Tr}f(\tilde P_I {\cal O}\tilde P_I{\cal O}^{-1})+(N_I-\tilde N_I)(f(1)-f(0))~,}
the first two terms in each of \rwe\ and \rwet\ are separately equal (modulo irrelevant constant terms) if $c_I=-\tilde c_I$. To establish the equivalence of the third terms we need an extra matrix identity. It is
\eqn\sdf{
{{\rm det} M_+\over{\rm det}\tilde M_+}={\rm det}~e^{-V}~.}
To prove it define the square matrix
\eqn\sazx{
\hat\Phi=\left(\matrix{{\bf 1}_{N_I\times N_I} & Q_{N_I\times\tilde N_I}\cr 0_{\tilde N_I\times N_I} & {\bf 1}_{\tilde N_I\times N_I}}\right)~,}
which has detrminant equal to one, and thus ${\rm det}~(\hat\Phi e^{-V}\bar{\hat\Phi})={\rm det}~ e^{-V}$. Now, for any invertible matrix ${\cal M}=\left(\matrix{A & B \cr C & D}\right)$ with inverse ${\cal M}^{-1}=\left(\matrix{X & Y \cr Z & W}\right)$ the following identity holds
 \eqn\pkj{{\rm det}~{\cal M}={{\rm det}~A\over{\rm det}~ W}.}
 Applying this to ${\cal M}=\hat\Phi e^{-V}\bar{\hat\Phi}=\left(\matrix{M_+ & *\cr * & *}\right)$,  ${\cal M}^{-1}=\left(\matrix{* & *\cr * & \tilde M_+}\right)$ one sees that \sdf\ holds. This entails that 
\eqn\tf{{\rm Tr}~{\rm ln}~M_+ = -{\rm Tr}~V+{\rm Tr}~{\rm ln}~\tilde M_+=-\sum_i a_{Ii}{\rm Tr}~V_i+{\rm Tr}~{\rm ln}~\tilde M_+~,}
and so the first three terms in \rwe\ and \rwet\ are equal provided $c_I=-\tilde c_I$ and $c_{i}\rightarrow\tilde c_{i}=c_{i}+a_{iI}c_i$. 

There is one final step to complete the proof. For this one has to look at the holomorphic constraints at the nodes connected to node $I$. These are
\eqn\newfterms{\eqalign{
- \sum_{m=1}^{a_{Ij}}\Phi_{jI}^m \Phi_{Ij}^m + \sum_{m=1}^{a_{ij}}\sum_{i\neq I}\Phi_{ji}^m \Phi_{ij}^m &= b_j~,\cr
- \sum_{m=1}^{a_{Ij}}\tilde\Phi_{jI}^m \tilde\Phi_{Ij}^m + \sum_{m=1}^{a_{ij}}\sum_{i\neq I}\Phi_{ji}^m \Phi_{ij}^m &= \tilde b_j~.
}}
which on taking traces and using \xxxxx\ yields $\tilde b_j=b_j +a_{Ij}b_I$, 
where $b_I=-\tilde b_I$ has been used. All in all, then, the dual FI parameters are related as
\eqn\fipshift{
\tilde c_i=c_i + (-2\delta_{iI}+a_{iI})c_I~,~~~~~~~~\tilde b_i= b_i + (-2\delta_{iI}+a_{iI})b_I~.}
Note that the shift is the one expected by the $SU(2)_R$ symmetry of $N=2$ supersymmetry.

A few comments are in order. First, the proof given above is classical: the hyper-K\"ahler quotient gives the moduli space on the Higgs branch of the gauge theory, where the gauge symmetry is completely broken. This is similar in spirit to \BerensteinFI, where the $N=1$ case was considered. \foot{In fact, our mapping between the dual variables bears a striking resemblance to theirs. We believe that direct application of our methods in the case of K\"ahler quotients should be enough to prove the equivalence at the level of K\"ahler potentials for the $N=1$ case, after enforcing the constraints imposed by the dual superpotential.} Second, the proof is applicable to {\it arbitrary} $N=2$ quiver theories, independently of their embedding in string theory. Also, one can incorporate massless fundamental flavors straightforwardly. All one has to is include the corresponding node with its label indicating the number of flavors, but freeze the gauge group and remove the FI parameter. The transformation of the node indices is the same, where now one includes the indices associated to the flavor nodes. The FI parameters of the gauged nodes transform also as before. Finally, we would like to point that, at the classical level that we are considering, $N=2$ Seiberg duality holds in the limit case in which $\sum_i a_{Ii}N_i=N_I$. In the simplest situation of $U(N_c)$ gauge theory with $N_f$ flavors this is precisely the case when $N_c=N_f$. We discuss it next.
 
\subsec{The limit case: $N_f=N_c$}
Our previous discussion makes it clear how to treat this extreme case at the classical level. When $\sum_i a_{Ii}N_i=N_I$ the fields $\Phi_{\pm}$ are $N_I\times N_I$ matrices. The gauge choice \soc\ then sets $\Phi_+$ to be the identity. Solving the F-term constraints then gives $\Phi_-=b_I{\bf 1}$. The matrices $M_+$ and $M_-$ then become proportional to the identity, and the terms of the hyper-K\"ahler potential coming from node $I$ give irrelevant constants. In the dual theory node $I$ has zero rank. The quotient can be performed removing this node and all the fields attached to it. This is clear, given that in the original theory all fields attached to $I$ could be gauged away. There are however some compatibility conditions between the new and old FI parameters at the nodes attached to $I$. Equation \tf\ gives $\tilde c_j=c_j+a_{Ij}c_I$. The rest are easily derived from equations \newfterms, which in this case give 
\eqn\newftermsbis{
- a_{Ij} b_I+ \sum_{m=1}^{a_{ij}}\sum_{i\neq I}\Phi_{ji}^m \Phi_{ij}^m = b_j,~~~~~ \sum_{m=1}^{a_{ij}}\sum_{i\neq I}\Phi_{ji}^m \Phi_{ij}^m = \tilde b_j~.
}
Taking traces in the previous equation we obtain $\tilde b_j=b_j+a_{Ij}b_I$.

The previous discussion has an interesting and curious application to $ALE$ spaces. We see that, apart as being the hyper-K\"ahler moduli spaces of extended $ADE$ Dynkin quivers, they can be obtained as hyper-K\"ahler quotients of quivers with extra nodes which can all be $N=2$ Seiberg dualized to zero. The previous compatibility amongst FI (resolution) parameters ensures that there are no additional deformation moduli associated to these new realizations. Next we discuss two simple examples of these ``exotic'' realizations of $ALE$ spaces. 

\fig{To the left the usual $\hat{A}_1$ quiver. Nodes are labeled from left to right as $0$, $1$. The corresponding gauge fields are $V_1$ and $V_2$. $w_{\pm}$, $t_{\pm}$ have charges $( \mp 1, \pm 1)$ under $(V_1 , V_2)$ . To the right is the extended quiver, where nodes are labelled as $V_1 , V_2 , V_3$ from left to right, again. The charges are $( \mp 1 , \pm 1 , 0)$ for $w_{\pm}$, $t_{\pm}$, and $(0, \mp 1 ,  \pm 1)$ for $z_{\pm}$.}{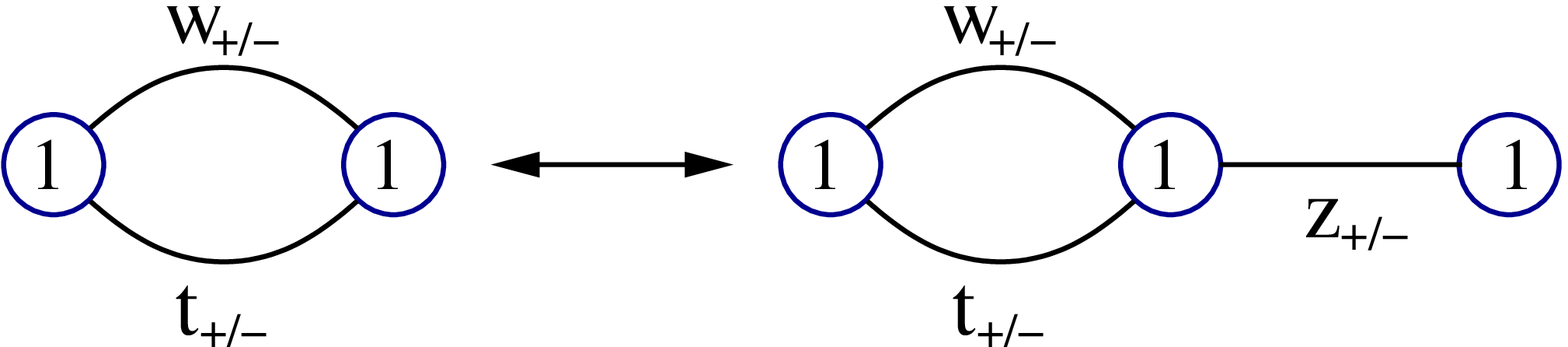}{115mm}

Consider the quivers in Fig. 2. It is known that the hyper-K\"ahler quotient of the quiver on the left gives the Eguchi-Hanson manifold \EguchiXP. The gauged K\"ahler potential of the right quiver is
\eqn\hkqo{\eqalign{
K=\int d^4\theta ~&\left[w_+ e^{V_1} \bar w_+ e^{-V_2} + w_- e^{-V_1} \bar w_- e^{V_2} +  t_+ e^{V_1} \bar t_+ e^{-V_2} + t_- e^{-V_1} \bar t_- e^{V_2} \right]+ \cr
& + \int d^4\theta~ \left[z_+ e^{V_2} \bar z_+ e^{-V_3} + z_-e^{-V_2} \bar z_- e^{V_3} - c_1~V_1 - c_2~V_2 -c_3~V_3~\right].
}}
This is subject to the F-term constraints
\eqn\fiseven{\eqalign{
w_{+}w_{-}+t_{+}t_{-}&=b_1 \cr
-w_{-}w_{+}-t_{-}t_{+}+z_{+}z_{-}&= b_2 \cr
-z_{-}z_{+}&=b_3~.
}}
The field equation for $V_3$ yields
\eqn\fefvt{
e^{V_3}=e^{V_2}{c_3+\sqrt{c_3^2+4(z_+\bar z_+ )(z_-\bar z_- )}\over 2z_-\bar z_-}~.}
The third equation in \fiseven\ implies that $z_+$ and $z_-$ are related as $z_+=b_3 / z_-$. Using the complex gauge invariance at node $3$ we can set $z_+=1$ and thus $z_-=b_3$. Modulo irrelevant constant terms, the gauged K\"ahler potential becomes
\eqn\gtfsd{\eqalign{
K=\int d^4\theta~&\left[w_+ e^{V_1} \bar w_+ e^{-V_2} + w_- e^{-V_1} \bar w_- e^{V_2} +  t_+ e^{V_1} \bar t_+ e^{-V_2} + t_- e^{-V_1} \bar t_- e^{V_2} \right]+ \cr
& + \int d^4\theta~\left[ -c_1~V_1 - (c_2+c_3)~ V_2\right]~,}}
subject now to
\eqn\neftrm{\eqalign{
w_{+}w_{-}+t_{+}t_{-}&=b_1 \cr
-w_{-}w_{+}-t_{-}t_{+}&= b_2+b_3 ~.}}
Now we can safely perform the rest of the hyper-K\"ahler quotient. Denoting $V=V_1 -V_2$ and $p=e^V$, and writing the equation of motion for $V$ we obtain
\eqn\pequal{
p={c_1 + \left[c_1^2+4(w_{+} \bar w_+ + t_+ \bar t_+ )( w_- \bar w_- + t_- \bar t_-)\right]^{1/2}\over 2(w_+ \bar w_+ + t_+ \bar t_+)}~,}
where $c_1 +c_2 +c_3=0$ has been used. Fixing the remaining gauge symmetry and imposing the F-term constraints  we find
\eqn\pis{
p={1\over 2(1+|z|^2)}\left[ c_1 + (c_1^2 + 4(1+|z|^2 )^2 |w|^2)^{1/2}\right]~,
}
where we have defined $z=t_+$and $w=t_-$. Finally, substituting back into $K$ we find \foot{We have used the $SU(2)_R$ symmetry to make the FI parameters real ($b_1 =b_2 =b_3=0$) to simplify the final expression.}
\eqn\sbi{
K=\int d^4\theta~ \left[ (c_1^{2}+4(1+|z|^{2})^{2}|w|^{2})^{1/2}-c_1~\log~p\right] ~,
}
which is the Eguchi-Hanson K\"ahler potential \CalabiAE\DouglasZJ.

Consider next the two quivers in Fig. 3. 
\fig{Another two equivalent quivers. The notation is the same as in Fig. 5, but now $z^{(i)}_{+}$ ($z^{i}_{-}$) is a doublet (antidoublet) under the $U(2)$ of the additional node.}{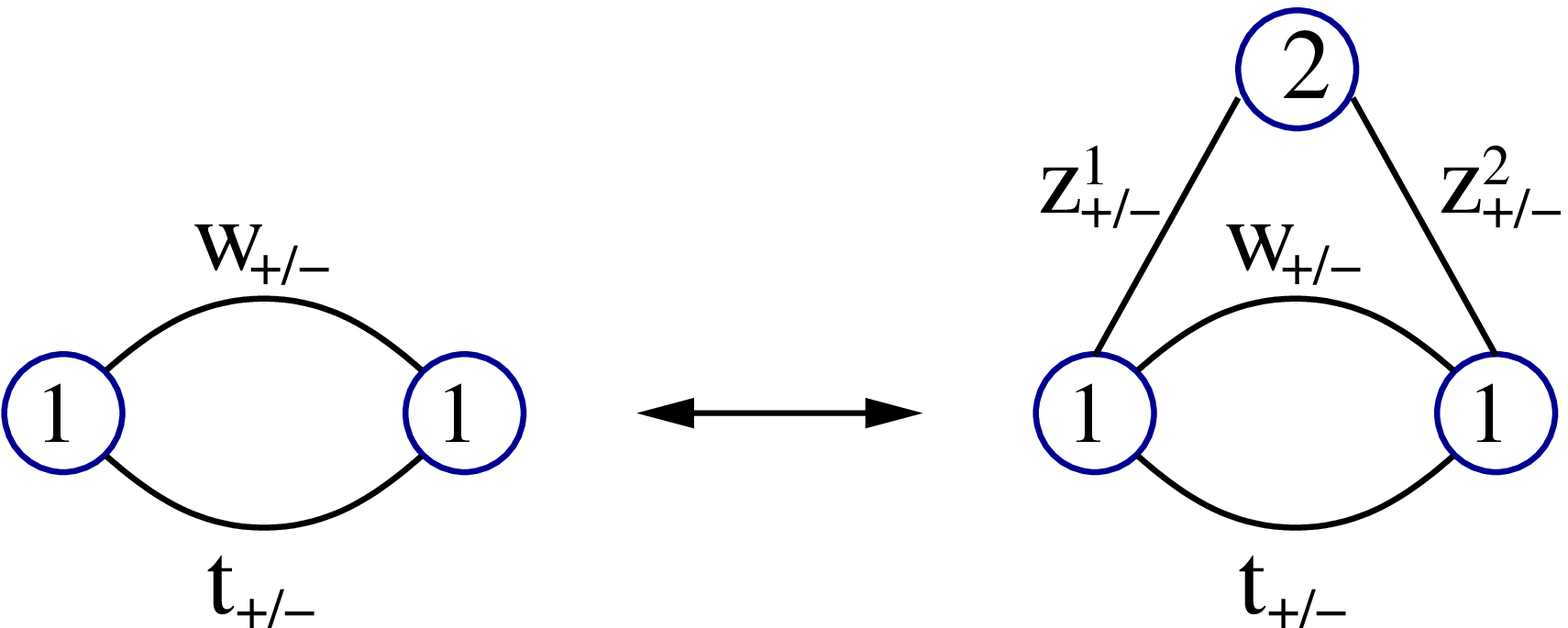}{100mm}
The gauged K\"ahler potential corresponding to the right quiver is
\eqn\wfwtgkp{\eqalign{
K=\int & d^4\theta~\left[{\rm Tr}\left(\bar z_+^1 e^{V_3} z_+^1 e^{-V_1}+ \bar z_-^1 e^{-V_3} z_1^1 e^{V_1} + \bar z_+^2 e^{V_3} z_+^2 e^{-V_2}+ \bar z_-^2 e^{-V_3} z_-^2 e^{V_2}\right)\right]+ \cr
 & +\int d^4\theta~ \left[ \bar w_+ w_+ e^{(V_{2}-V_{1})}+ \bar w_- w_- e^{-(V_{2}-V_{1})}+ \bar t _+ t_+ e^{(V_{2}-V_{1})} + \bar t_- t_- e^{-(V_{2}-V_{1})}\right]+ \cr
& +\int d^4\theta~ \left[- c_1~V_1 - c_2~ V_2 - c_3 ~{\rm TrV_{3}}\right]~.
}}
subject to the F-term constraints\foot{To simplify the discussion, once again we have made use of the $SU(2)_R$ symmetry to rotate the FI parameters.}
\eqn\atikp{\eqalign{
w_{+}w_{-}+t_{+}t_{-}+z_{+}^{1 \alpha}z_{-\alpha}^1&=0 \cr
 -w_{-}w_{+}-t_{-}t_{+}+z_{+}^{2\alpha}z^{2}_{-\alpha}&=0 \cr
z_{-\alpha}^{1}z_{+}^{1 \beta}+z_{-\alpha}^{2}z_{+}^{2\beta}&=0~.
}}
In this case, because there are two links connected to the would-be dualized node, we define
\eqn\bbgokl{
Z_+=\left(\matrix{z_+^{(1)1} & z_+^{(2)1}\cr z_+^{(1)2} & z_+^{(2)2}}\right) ~~~~~~~~~Z_-=\left(\matrix{z_{-1}^{(1)} &z_{-1}^{(2)} \cr z_{-2}^{(1)}  & z_{-2}^{(2)}}\right)}
and the diagonal matrix
\eqn\dgf{
V=\left( \matrix{V_1 & 0\cr 0 & V_2}\right)~,}
in terms of which the first line of \wfwtgkp\ can be written as
\eqn\flowk{
\int d^4\theta~\left[{\rm Tr}\left(\bar Z_+ e^{V_3} Z_+ e^{-V} +\bar Z_- e^V Z_- e^{-V_3}\right)\right]~.}
The equations of motion for $V_3$ now yield
\eqn\soleqmfvtr{
e^{V_3}={1\over 2}(Z_+ e^{-V}\bar Z_+)^{-1}\left(c_3+\sqrt{c_3^2+4(Z_+ e^{-V}\bar Z_+)(\bar Z_- e^{V} Z_-)}\right)~.}
Using the gauge symmetry at node $3$ we can fix $Z_+={\bf 1}$ and then through the last equation in the F-flatness conditions $Z_-=0$. Plugging this back into the gauged K\"ahler potential we get
\eqn\frd{\eqalign{
K=\int d^4\theta~ &\left[ \bar w_+ w_+ e^{(V_{1}-V_{0})}+ \bar w_- w_- e^{-(V_{1}-V_{0})}+ \bar t _+ t_+ e^{(V_{1}-V_{0})} + \bar t_- t_- e^{-(V_{1}-V_{0})}\right]+ \cr
& +\int d^4\theta~ \left[- (c_1+c_3)~V_1 - (c_2+c_3)~ V_2 \right]~.}}
We can now follow the steps of the previous example to get an Eguchi-Hanson manifold with deformation parameter $c_1+c_3$.

We now change gears to illustrate our results in the particular cases in which the $N=2$ theories can be embedded in string theory. On order to do this we first briefly review the most popular constructions of gauge theories from string theory.

\newsec{$N=2$ gauge theories from string theory}

\subsec{Geometric engineering of $N=2$ $ADE$ gauge theories}
It is well known that one can engineer four-dimensional gauge theories with $N=2$ supersymmetry starting with type $IIA$ or $IIB$ string theory ``compactified'' on singular Calabi-Yau's which are  $K3$ fibrations over ${\bf CP^1}$ \KatzEQ\CachazoSG. To decouple gravitational and stringy effects and recover enhanced $ADE$ gauge symmetry the $K3$ fiber must degenerate to its $ADE$ singular limit and the ${\bf CP^1}$ base must have infinite volume. When one wraps $D$ branes around the vanishing cycles of the geometry new degrees of freedom arise from the strings connecting the $D$-branes, which become massless whenever the two wrapped on cycles intersect. For concreteness let us review the $IIB$ setup \CachazoSG\ and see this in more detail. Consider a set of $D3$ branes transverse to a singular six dimensional space described by an equation 
\eqn\locsing{
f(x,y,z)=0~,}
in $(x,y,z,w)$, where $f(x,y,z)$ is the equation for an $ADE$ singularity (see \tthree), and $w$ is the coordinate of a complex plane. In the neighborhood of the singularity this is an ${\cal O}(0)\oplus{\cal O}(-2)$ bundle over ${\bf CP}^1$. The four dimensional gauge theory in the worldvolume of the $D3$ branes has eight supercharges, or $N=2$ supersymmetry in four dimensions.

As found in \DouglasSW\ (see also \JohnsonPY) the worldvolume action is found by orbifold projecting the ten dimensional super Yang-Mills action under the discrete group associated to the singularity, $\Gamma$, and its field content can be summarized in an associated (affine) $ADE$ Dynkin diagram (quiver), which we denote as $\Delta(\Gamma)$. Moreover twisted fields from the closed string sector couple as FI parameters in the gauge theory, through the Chern-Simons term in the $D$ brane action $\sum_k \int C_k\wedge{\rm Tr}~e^{(F-B)}$. In particular the Ramond-Ramond form $C_k$ from the $k$-th twisted sector couples to the $U(1)$ part of the gauge field of the $D$ brane whose Chan Paton factor is acted by the $k$-th irreducible representation. In addition to the transverse $D3$ branes, one can wrap $N_i$ $D5$ branes around the vanishing ${\bf CP}^1_k$ of the geometry. These are fractional branes \DiaconescuBR. Altogether the theory with $N$ $D3$ branes and $N_k$ $D5$ branes wrapped around ${\bf CP}^1_k$ has gauge group $G=\prod_k U(N+N_k)$, with $k$ running over the nodes of $\Delta(\Gamma)$. It also has $a_{ij}$ bifundamental hypermultiplets $(\Phi_{ij},\Phi_{ji})$ between nodes $i$ and $j$, with $a_{ij}$ the adjacency matrix of $\Delta(\Gamma)$. These theories have the lagrangian
\eqn\twentynine{\eqalign{
{\cal L}=\int d^4 &\theta ~ \left[\sum_i{\rm Tr}\left(\bar S_ie^{V_i}S_i\right)+\sum_{i\neq j}a_{ij} {\rm Tr} \left( \Phi_{ij}e^{-V_j}\bar\Phi_{ij}e^{V_i}\right) ~-~\sum_{i}c_i{\rm Tr}~V_i\right] \cr
+&\int d^2 \theta \left[ \sum_{i,j} s_{ij} {\rm Tr} \left(\Phi_{ij} S_j \Phi_{ji}\right) - \sum_i b_i {\rm Tr}~ S_i \right]~+({\rm h.c.}),
}}
with $b_i,c_i,\bar b_i$ the triplet of $N=2$ FI parameters \foot{In the context of string theory, the $FI$ parameters correspond to the sizes of the blown-up spheres and control the masses of the gauge bosons arising from $D5$-branes wrapping the vanishing cycles. Moreover, the K\"ahler class of the manifold is complexified due to the presence of the $B$-fields. In type $IIB$ one can switch on $B_{RR}$ and $B_{NSNS}$ thereby changing the stringy K\"ahler volume of the spheres to
\eqn\twentyeight{
V_{S^2_i} = \left[ |{B_{NSNS}}_i + i{B_{RR}}i|^2 +  |c_i + iB_i|^2\right]^{1/2}~,
}
and the $i$-th group gauge coupling constant to $g_i^{-2}=V_{S^2_i}/g_s$, where $g_s$ is the $IIB$ string coupling.} and $s_{ij}=-s_{ji}$ equal to $\pm1$ if nodes $i$ and $j$ are connected and zero otherwise.
The moduli space is the set of gauge invariant solutions of
\eqn\thirty{\eqalign{
[S_i,\bar S_i]&=0~,\cr
a_{ij} \left( \Phi_{ij} \bar\Phi_{ij} - \bar\Phi_{ji} \Phi_{ji}\right) &= c_i {\bf 1}~,\cr
\sum_j \left(s_{ij} \Phi_{ij} \Phi_{ji}\right) &= b_i {\bf 1} ~,\cr
S_i \Phi_{ji} - \Phi_{ij} S_j &= 0~.
}}
This moduli space contains various branches: baryonic and non-baryonic Higgs branches, Coulomb branch, and mixed branches. In the pure Higgs branch where all $S_i$ are zero and $\Phi$'s have maximal rank
 one is left with the second and third equations above. The solution to these is precisely the moduli space of the corresponding quiver, as in Section 2. For completeness the representation theory of the algebra corresponding to Eqs. \thirty\ is reviewed in the Appendix.

\subsec{Generalized Hanany-Witten setups}
Generalized Hanany-Witten setups \HananyIE\WittenSC\ (see also \OhBF) provide another useful perspective in constructing $N=2$ gauge theories with bifundamental matter. They arise naturally after performing T-duality on the previous setup. To see this start with a geometrically engineered $N=2$ theory as in the previous subsection.\foot{For simplicity we give details for the $A_{n-1}$ case only.} The geometry is locally given by 
\eqn\geom{
xy+z^{n}=0~.}
This is singular at $x=y=z=0$. To resolve the singularity consider patches coordinatized by $(x_i,y_i)$, $i=1,\ldots,n$ subject to
\eqn\resanmo{\eqalign{
x_n&=x,~~x_{n-1}={x_n / z}={x/ z},~~\ldots,~~x_{n-i}={x_{n-i+1}/ z}={x/ z^i}~\cr
y_1&=y,~~y_2={y_1/ z}={y/ z},~~\ldots,~~y_n={y_{n-1}/ z}={y/ z^{n-1}}~.}}
One has
\eqn\polic{
x_i y_i=xy/z^{n-1}=z~,}
so in each of these patches the equation is non-singular. The patches are glued together as
\eqn\gluing{
x_i y_i =x_{i+1}y_{i+1},~~~~x_i y_{i+1}=1~.}
The blow-down map is straightforward
\eqn\blowdown{
x=x_i^{n-i+1}y_i^{n-1},~~~~~y=y_i^i x_i^{i-1},~~~~~z=x_i y_i~.}
It is easy to see that the resolved geometry contains $n$ ${\bf CP}^1$'s parametrized by $x_{i+1}=y_i=0$, which intersect at $x_i=y_i=0$. Consider the circle action $(e^{i\theta},x_i)=e^{i\theta}x_i$ and $(e^{i\theta},y_i)=e^{-i\theta}y_i$, which is compatible with the gluing \gluing\ and thus can be extended globally. It leaves the ${\bf CP}^1_i$'s invariant. Performing T-duality along the angular direction turns the $A_{n-1}$ singularity into $n$ NS5-branes located at $x_i=y_i=0$ \OoguriWJ\UrangaVF\OhBF (where the circle action degenerates) and turns the wrapped D5-branes into D4-branes stretching across consecutive NS5-branes. This is illustrated graphically in Fig. 4.\fig{The relation between geometrically engineered $N=2$ quiver gauge theories. On the right NS5-branes are depicted as vertical lines and D4-branes as horizontal lines.}{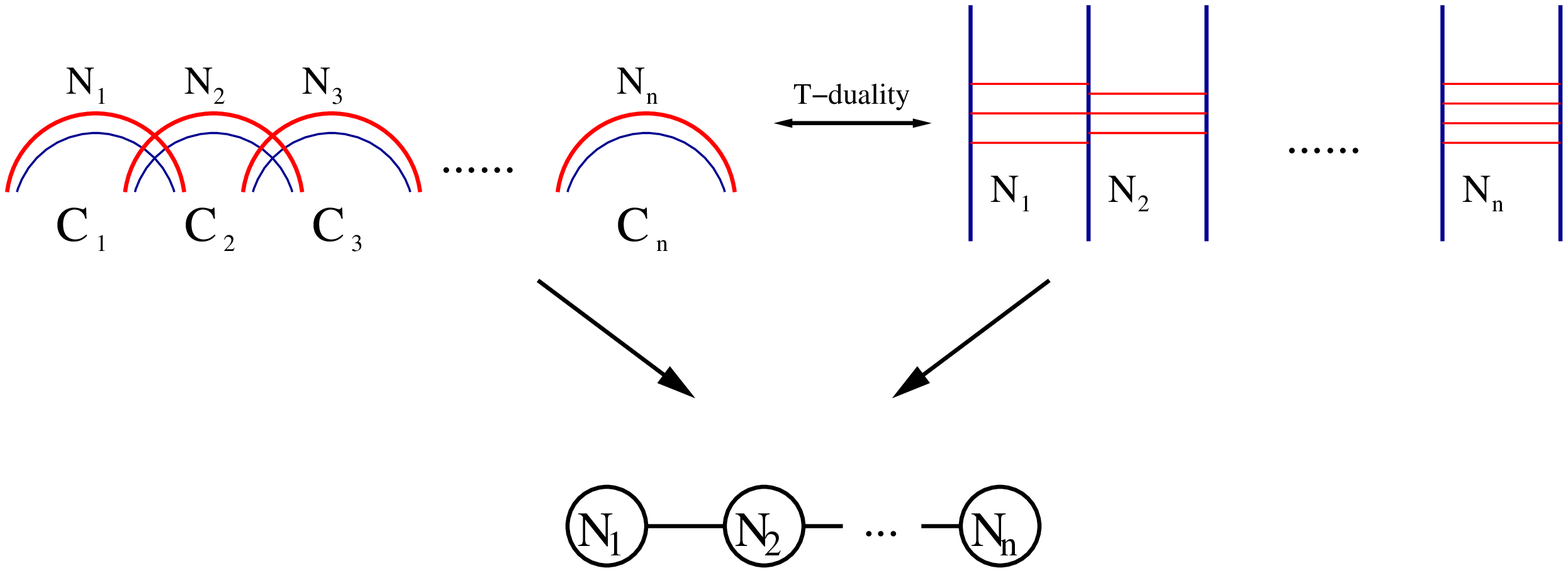}{140mm}

In the conventional choice for coordinates, at the classical level the $n$ $NS5$ branes fill the $012345$ directions and are located at fixed positions in $6789$; between any two consecutive $NS5$ branes we have $N_k$, $k=1,\ldots, n$ parallel $D4$-branes filling $01236$. The $D4$-branes are not infinite in the $6$ direction, but end on the $(k-1)$-th and $k$-th $NS5$ branes from the left and right respectively. The classical interpretation of different fields and parameters is as follows:

-the coupling constant of the $U(N_k)$ factor is $g_k^{-2}={\Delta x^6_i \over g_s}$, where $\Delta x^6_k$ is the separation of the $k-1$-th and $k$-th $NS5$ branes in the sixth direction, and $g_{s}$ is the $IIA$ string coupling constant.

-the motion of the $D4$-branes along $45$ is parameterized by a chiral multiplet and gives the (classical) Coulomb branch of the gauge theory. Together with the gauge field in the worldvolume it contitutes an $N=2$ vectormultiplet.

-the separation of the $NS5$-branes in $789$ play the role of $N=2$ FI parameters of the gauge theory.

-there are bifundamental hypermultiplets arising from strings stretching between adjacent $D4$-branes. These become massless whenever the $D4$-branes become coincident in $45$. In the affine case (elliptic setups) these hypermultiplets can acquire VEVs, which parameterize the motion if the $D4$ branes in $789$, or the Higgs branch of the theory.

In the non-elliptic case (ordinary, non affine $ADE$ quivers) for this configuration to be possible the $NS5$-branes must be coincident in $789$. This means that the $D4$ branes cannot move in the $789$ space, as they are constrained to end on the $NS5$ branes. Moreover, we saw above that when all the $NS5$ branes are located at the same point in $789$ their transverse space is $T$-dual to $A_{n-1}$ singularity. In other words, the $D4$ branes are stuck at the orbifold singularity, and the Higgs branch has either an isolated vacuum (when complete Higgsing is possible, and the associated quiver gives a zero dimensional space through the hyper-K\"ahler quotient procedure) or is completely lifted (when the gauge groups are such that no complete Higgsing is possible, and therefore the hyper-K\"ahler quotient derived from the associated quiver diagram would give a negative dimension for the quotient). In the terminology of the Appendix, these zero dimensional Higgs branches correspond to positive roots of the corresponding $ADE$ group.

In the elliptic setups the $6$ direction is periodic. Alternatively, the associated $ADE$ algebra is of affine type. If all the $N_{k}=N$ are equal, then the $D4$-branes can reconnect across $NS5$-branes and split off them as $N$ $D4$-branes wrapping the $6$ dimension. Therefore they are no longer constrained to end on the $NS5$-branes and can freely move in the $789$ directions. As a consequence, one is now free to move the $NS5$-branes apart from one another in the $789$ directions. This splitting corresponds to switching on the $FI$ parameters. This resolves the $T$-dual $A_{n-1}$ singularity and one recovers the smooth $ALE$ space in the dual picture. These Higgs branches are exact classically (the gauge symmetry is fully broken),  and they can be computed using the hyper-K\"ahler quotient procedure. Again, in the terminology of the Appendix, these are null roots of the affine $ADE$ algebra.

On the other hand, for a Coulomb branch to exist, we saw that the $NS5$ branes must be coincident in $789$. This means that the $FI$ parameters must be zero, as one would expect from gauge theory considerations. Classically these Coulomb branches are parameterized by the independent motion of the ${\sum}N_{k}$ $D4$ branes in the $45$ space. However, they receive perturbative and non-perturbative corrections. Perturbatively, these corrections arise from the bending of $NS5$-branes due to the $D4$-branes ending on them. In this way the stretching of the $D4$ branes in the $6$ direction is corrected because of the non-trivial shape of the $NS5$-branes, and one recovers the perturbative running of the $N=2$ coupling constants in four dimensions. Moreover, there is an anomaly whose cancellation freezes the $U(1)$ factors of the gauge groups. To fully solve the theory, one can lift the whole configuration to $M$-theory, where the $NS5$ and $D4$ branes are seen to arise from an $M5$-brane wrapping the Seiberg-Witten curve \WittenSC. We will not discuss it here.

Finally, and this will be important for us, one can also incorporate $M_k$ $D6$ branes filling $0123789$, and located at points in the $x^{6}$ which are between the location of the $(k-1)$-th and $k$-th $NS5$-branes \WittenSC. Strings stretching between the $N_{k}$ $D4$ branes and the $M_{k}$ $D6$ branes give rise to $M_{k}$ hypermultiplets in the fundamental representation of $U(N_{k})$. These strings will become massless whenever the $D4$ branes and the $D6$ branes become coincident in $45$.

Having now reviewed this type of constructions, we can give a very intuitive picture of our construction in Section 3.

\newsec{$N=2$ Seiberg duality in string theory}

The kind of Seiberg duality for $N=2$ theories we discussed in Section 3 has a natural interpretation in string theory, which goes around very similar lines as the original picture of \ElitzurHC. Although the discussion in Section 3 was for general $N=2$ gauge theories in the Higgs branch with an arbitrary number of links connected to a node, one can see that the main features appear naturally from brane constructions of gauge theories which can be engineered from string theory. We consider several cases in turn.

\subsec{$U(N)$ with $M$ flavors}
 \fig{Brane configuration for a $U(N)$ gauge theory with $M$ in the baryonic Higgs branch, for $N=2$, $M=5$. $NS5$ branes are depicted as continuous vertical lines, $D4$ branes as horizontal lines, and $D6$ branes as vertical dashed lines.}{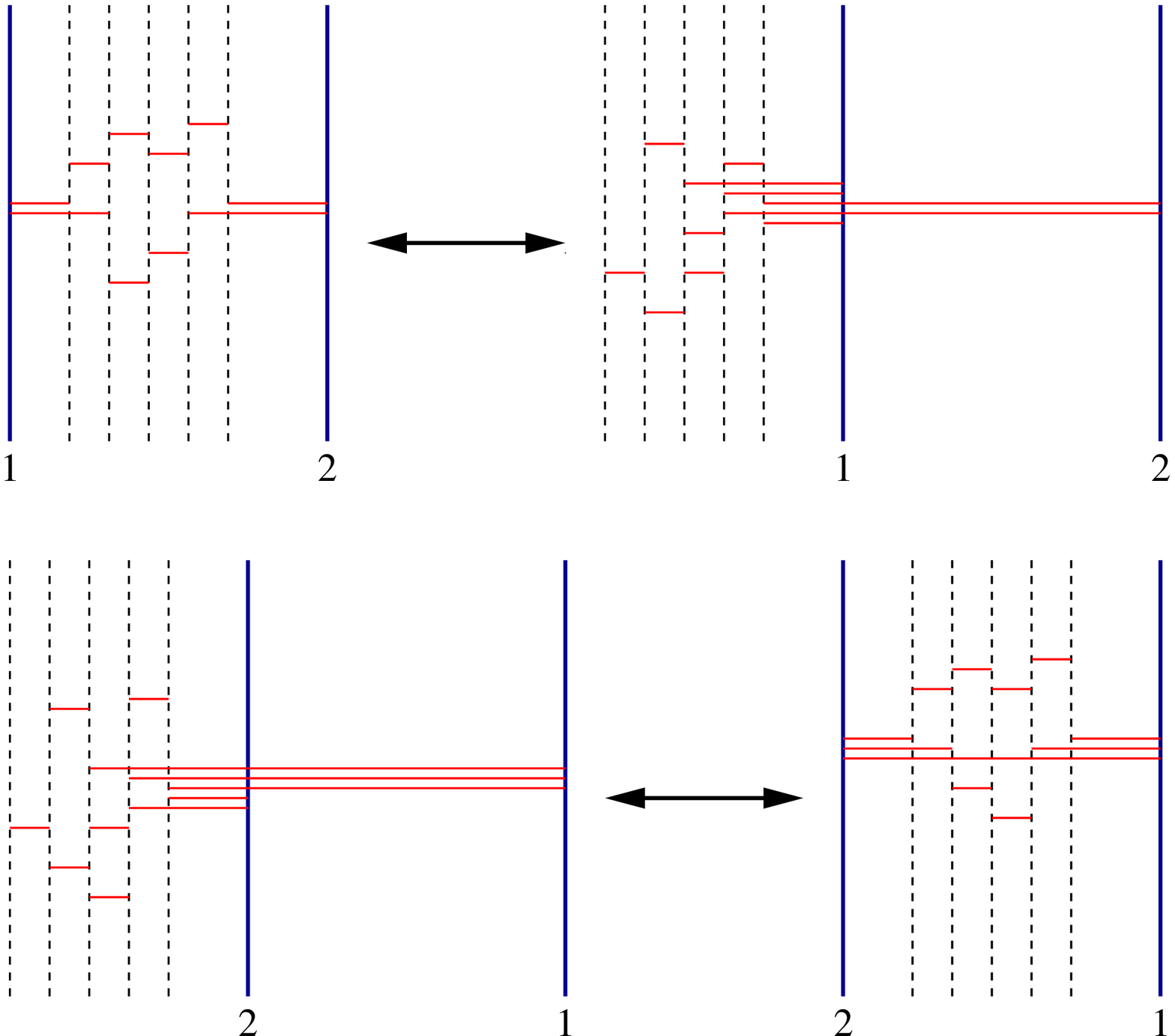}{90mm} 
Take a theory with $U(N)$ gauge group and $M$ hypermultiplet flavors in the Higgs branch. This is illustrated in the Hanany-Witten setup in Fig. 5 (upper left). In the initial configuration flavor hypermultiplets arise through $D4-D6$ and $D6-D4$ strings, which become massless when the $D4$ and $D6$ branes become coincident in $45$. In the baryonic Higgs branch the gauge symmetry is completely broken. The $D4$ branes break up on $D6$ branes in the most general way consistent with the $s$-rule, so that no more than one $D4$ brane stretches between the same $NS5$- $D6$ brane pair \HananyIE. The remaining segments of the $D4$ branes move in the $789$ directions and their positions, together with the Wilson line in the $6$ direction, parameterize the Higgs branch of the gauge theory. One can give an alternative descrition of the Higgs branch as follows. The $x^6$ position of the $D6$ branes does not play a role in the low-energy theory \WittenSC. We can then move them to the left of the left $NS5$ brane. Through the Hanany-Witten effect a new $D4$ brane is created for each of the $D6$ branes crossing the $NS5$. Massless hypermultiplets arise now through $D4-D4$ strings stretching across the $NS5$ brane. Note that the newly created $D4$ branes have no moduli because of the boundary conditions at each end. Their gauge symmetry is frozen, and becomes effectively a flavor symmetry. We have then $M$ chiral multiplets in the $N$ representation of the gauge group and $M$ chiral multiplets transforming in the $\bar N$ coming from oppositely oriented strings. As before, the $D4$ branes can break on $D6$ branes and the system undergoes a transition to the Higgs branch. This can happen, and correctly reproduce the dimension of the Higgs branch without violating the $s$-rule, because the $N$ $D4$ branes stretching between the two $NS5$ branes can reconnect with $N$ of the newly created $D4$ branes. The baryonic Higgs branch is depicted in Fig. 5 (upper right). By displacing the left $NS5$ brane in the $789$ directions (as now there are only $D4$ branes connecting it to the $D6$ branes, whose worldvolume extends in the $789$ directions) one can move the left $NS5$ brane to the right of the other $NS5$ brane and avoid the singular situation in which both $NS5$ branes coincide in spacetime. In the gauge theory this is achieved by turning on the FI parameter of $U(1)\in U(N)$. We finally have the situation depicted in Fig. 5 (lower right), which corresponds to a $U(M-N)$ gauge group and $M$ flavors. Note that the ordering of the $NS5$ branes is the opposite of the one we started with. The segments of $D4$ branes are now moving in the transverse space to the $NS5$ branes, which can be related to the initial transverse space by changing the sign of the FI parameters. 
In quiver language, one started with the $A_2$ quiver in the configuration $Ne_1+Me_2$. The Higgs branch of the $U(N)$ gauge theory with $M$ flavors is just the hyper-K\"ahler quotient manifold associated to this quiver, taking into account that the node $e_2$ has no gauge symmetry associated to it. However this does not affect the duality discussed in Section 3. By performing the change of basis $e_1\rightarrow -e_1$ (interchanging the $NS5$ branes) this is the same configuration as $(M-N)(-e_1)+M(e_1+e_2)$. 

\subsec{Mixed theories}

\fig{$N=2$ Seiberg duality with colors and flavors.}{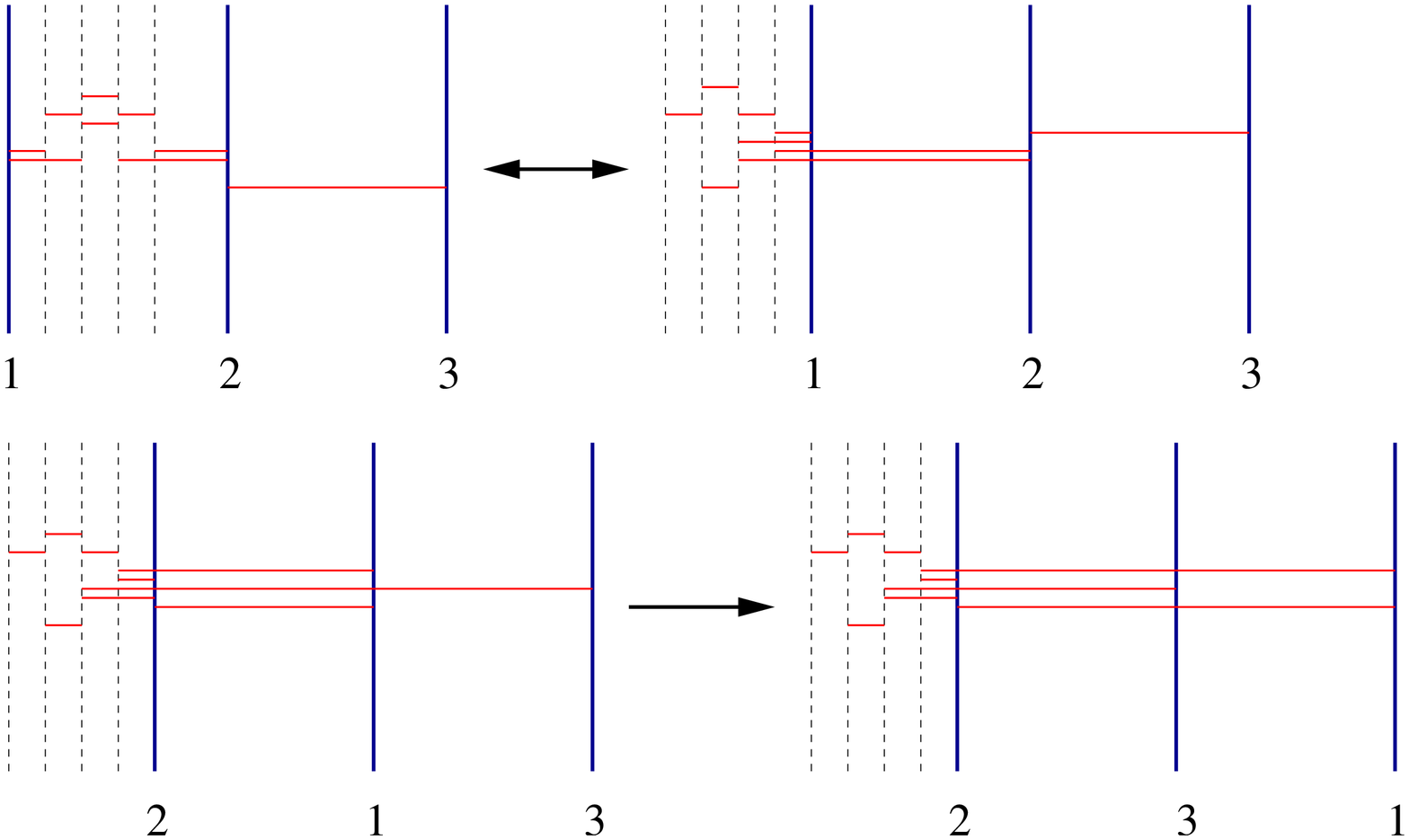}{110mm}
Using again the results of Section 3, we can easily derive a set of $N=2$ Seiberg-like dualities involving both colors and flavors. Consider the setup in Fig. 6. On the upper left is depicted the Higgs branch of a theory with gauge group $U(N_1)\times U(N_2)$ and $M_1$ massless flavors \foot{We are taking the $D6$ branes to be coincident with the $D4$ branes in the $45$ directions.} transforming in the fundamental representation of $U(N_1)$, in the particular case in which $N_1=3$, $M_1=4$, $N_2=1$. In the upper right of the figure is the alternative description in which all $D6$ branes have been pushed to the left of the $NS5$ branes. By applying similar steps to the ones in the previous subsection we arrive at the baryonic Higgs branch of a $U(M_1+N_2-N_1)\times U(N_2)$ gauge theory with $M_1$ flavors in the fundamental representation of the first gauge group (lower left). Note again that one can move the first $NS5$ brane to the right of the second $NS5$ brane because there are no $D4$ branes stretched between them (two of them stretch between $D6$ branes and the second $NS5$ brane, and another two stretch between $D6$ branes and the first $NS5$ brane; one can them displace the first $NS5$ brane in the $789$ directions, thus avoiding meeting the second $NS5$ brane in spacetime). If $N_1\geq N_2$, one can now apply the duality to the second gauge group. This is possible because all $D4$ branes can reconnect in such a way that there are no $D4$ branes stretching between $NS5$ branes $1$ and $3$. If there were such $D4$ branes one could not displace the $NS5$ brane in the $789$ directions without breaking supersymmetry. One gets to $U(M_1+N_2-N_1)\times U(M_1-N_1)$ gauge theory with $M$ flavors in the fundamental of the first gauge group. Once again brane motions exactly parallel the transformations of the FI parameters of Section 3.

It is clear that one can generalize these considerations to an arbitrary number of $NS5$ branes with $D4$ branes stretched between pairs. The construction of Section 3 guarantees that there is a large class of $N=2$ theories whose hyper-K\"ahler potential in the baryonic Higgs branch coincide, and that one can interpolate between them using brane motions akin to the ones just described. 

\newsec{A relationship with Fourier-Mukai transforms}
We would like to end this work with a curious interpretation of the transition in Fig. 2 \foot{The author is grateful to C. Vafa for suggesting this interpretation} in terms of Fourier-Mukai transforms for $K3$ \MukaiMF\HoriIQ. We will use the results reviewed in the Appendix. 

Take the $\hat A_1$ quiver theory, and consider the representation given by the null root $\delta$. The moduli space is simply the hyper-K\"ahler quotient of the $\hat A_1$ quiver, the Eguchi-Hanson manifold. This can be realized in a geometric engineering setup with a single $D3$ brane transverse to the $\hat A_1$ singularity. From this perspective, the moduli space is the resolved space probed by the $D3$ brane. Now consider an alternative realization: take a $D3-\bar{D5}$-boundstate wrapped around the cycle associated to $e_{0}$, and a $D5$-brane wrapped around $e_{1}$ (as in the transverse space a $D3$ brane wraps a 0-cycle, and a $D5$ brane wraps a 2-cycle, we will, following the terminology of \HoriIQ, call them $D0$ and $D2$ branes respectively). As made more explicit in the Appendix, this is a possible configuration of the $\hat A_1$ quiver, with root vectors given by
\eqn\fmt{\eqalign{
[D0-{\bar{D2}}]&=(e_{0}+e_{1})-(e_{1})=e_{0} \cr
[D2]&=e_{1}~.
}}
This simply means that the $H_{2}$ class of the bound state $[D0-D2]$ is given by $e_{0}$ and the one of $D2$ is simply $e_{1}$. This representation is of the form $N{\delta}$ (with $N=1$) and thus corresponds to an imaginary root of ${\hat{A}}_{1}$. Therefore the moduli space is the same as before.

Now let us perform $T$-duality on the transverse directions. The resolved $ALE$ space is non-compact in 3 directions, so $T$-duality per se does not make much sense. However, we can view the transverse $ADE$ singularity as a singular limit of $K3$. We can then study how the general $T$-duality of $K3$ affects the relevant cycles. It is known that such $T$-duality corresponds to a Fourier-Mukai transform, and in order to implement it according to the rules of \HoriIQ\ we need to know the charges of the $D$-branes present in our configuration.

A general configuration of $D$-branes (partially) wrapped around cycles in a manifold $X$ carries $RR$ charges specified by its Mukai vector \HoriIQ
\eqn\fmtch{
Q=v(E)={\rm ch}(E) \sqrt{\hat A (X)} = ({\rm rk}(E),{\rm c}_{1}(E),{1\over 2}{\rm c}_{1}^{2}(E) - {{\rm p}_{1}(X)\over 48}{\rm rk(E)})
}
where $E=F-B$, with $F$ the field strength of the worldvolume gauge theory. Also, $c_1(E)$ denotes the first Chern class of the bundle $E$, and $p_1(X)$ is the first Pontrjagin class. In writing the last equality we have expanded the Chern character of the bundle and the $A$-roof genus of the manifold and expressed the Mukai vector in a basis for the integral cohomology of $X$ given by
\eqn\fmbtr{
(H^{0}(X,{\bf{Z}}),H^{2}(X,{\bf{Z}}),H^{4}(X,{\bf{Z}}))~.
}
Now the Fourier-Mukai transform of the bundle acts as
\eqn\otrvlm{
rk(\hat{E})=c_{2}(E)-rk(E),~~~~~~
c_{2}(\hat{E})=c_{2}(E)=1
}
The Mukai vector of our initial $[D0-{\hat{D2}}],[D2]$ configuration is $(0,0,-1)$ (we conventionally take the D0-brane charge as -1). This corresponds to $rk(E)=0,c_{1}(E)=0,c_{2}(E)=1$. Therefore $rk(\hat{E})=1,c_{1}=0,c_{2}=1$, giving a transformed Mukai vector
\eqn\fffggg{
v(\hat{E})=(1,0,0)
}
where one has to take into account the curvature of $K3$ giving ${-p_1 (K3) \over 48}=1$. Using the description given in the Appendix this corresponds to a $D0-D4$ boundstate, with the initial $D2$ ${\bar{D2}}$ left unchanged as their total Chern charge adds to zero.

We need to know a little bit more about the integral cohomology of $K3$ to interpret this result in terms of the quiver diagram. As explained in {\it e. g.}
\lref\AspinwallMN{
P.~S.~Aspinwall,
arXiv:hep-th/9611137.
}\AspinwallMN\ the total cohomology lattice of $K3$ decomposes as
\eqn\kthree{
\Gamma_{4,20}=H^{*}(K3,{\bf Z})=H^{0}(K3,{\bf Z}){\oplus}H^{2}(K3,{\bf Z}){\oplus}H^{4}(K3,{\bf Z})={\Gamma}_{3,19}{\oplus}{\Gamma}_{1,1}
}
It is an even self-dual lattice of rank 24, with $(4,20)$ signature.The $\Gamma_{1,1}$ part gives the intersection matrix between 0- and 4-cycles. We can interpret the situation as follows: a new $D4$-brane is associated with the extra nodeof the exotic quiver, and intersects only the $D0$-brane according to ${\Gamma}_{1,1}$, and not the $D2$ or ${\bar{D2}}$-branes, giving the extra link of the quiver in Fig. 7. 

\fig{An exotic quiver associated to the $\hat A_1$ $ALE$ space.}{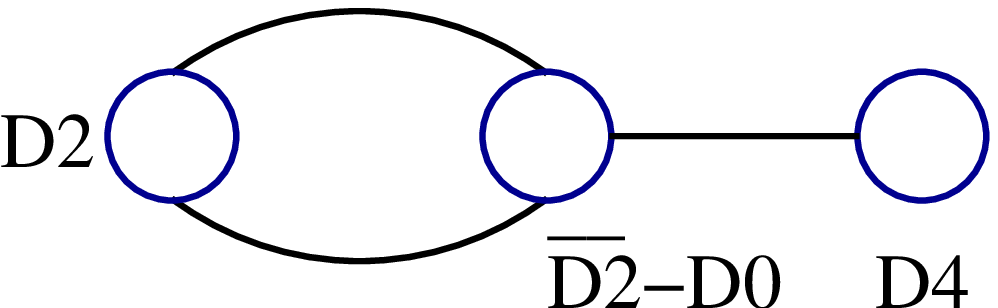}{65mm}

\vskip 0.5 cm{\bf Acknowledgements}

\vskip 0.5cm
The author wishes to thank Martin Ro\v cek for many useful discussions and for collaboration on related subjects.He aslo wishes to acknowledge many insightful remarks from Riccardo Ricci.This work was partially supported through NSF Grant No. 0098527.   

\appendix{A}{$ADE$ quiver representations}
A representation of a quiver \GabrielUD \KacIR\ is given by a complex vector space of dimension $N_k$ associated to node $k$, and a linear map $\Phi_{kl}~:~{\bf C}^{N_k}\longrightarrow {\bf C}^{N_l}$ to each arrow. The dimension vector of the representation is a vector $\vec v =(N_1 , \ldots ,N_r)$ specifying the dimension of the vector space associated to each node. In the $ADE$ case a theorem by Kac \KacIR\ states that the complex dimension of the moduli space of a quiver representation with dimension vector $\vec v$, given by $D=2-v^{T}Cv$ ($C$ is the Cartan matrix) is non-negative only for real positive roots of $G$ (null dimension) and for imaginary roots of $\hat G$ (positive dimension), with $G$ the corresponding $ADE$ group and $\hat G$ its affine extension. Note that this is what one expects from the hyper-K\"ahler quotient construction. 

Following \DouglasQW \FiolWX, let us be more specific the representration theory of $ADE$ quivers. If we choose a basis for the simple roots of $\hat G$, $\{e_1,\ldots,e_r\}$, the one for $\hat G$ can be seen as the extension 
\eqn\thirtythree{
\{ e_0 ,e_1 , \ldots, e_r \}~,
}
where the affine root is $e_0$ and $e_k \cdot e_l = a_{kl}$. A vector in the affine lattice can then be written as $(N,\vec v)$, where $\vec v$ belongs to the unextended lattice and $N$ is an integer. The root vectors of the non-affine $ADE$ are $(0,e_k)$ in the extended lattice, and $e_0 = (- \sum_{i=1}^r d_i e_i, 1)$, where $d_i$ are the indices of the Dynkin diagram. The lattice of positive roots is defined as
\eqn\thirtyfour{
\Gamma_+ = \{ k_1 e_1 +\ldots + k_r e_r + k_0 e_0 ~|~ k_i \geq 0 \}\,.}
These representations can be understood from the $D$ brane perspective as follows:

1) {\it positive roots of ADE}: ~these are $(\Delta^+ , 0 )$, where $\Delta^+ =\{ \sum_{i=1}^{r} N^k_i e_i \}$. To each positive root $\rho^k = n^k_1 e_1 +\ldots +n^k_r e_r \in \Delta^+$ is associated an irreducible representation of the algebra, labeled by the vector $(n^{k}_{1}, \ldots ,n^{k}_{r})$. They are interpreted as $D5$-branes wrapping the positive 2-cycle associated to $\rho^k$ in the homology lattice of $ALE$. These are fractional branes \DiaconescuBR, and are stuck at the singularity. In other words, the hyper-K\"ahler quotient aoosiacted to these representations gives a negative dimension for the moduli space. There is no Higgs branch of the corresponding quiver gauge theory.One can combine several of these representations as
\eqn\thirtyfive{
M_1 \rho^1 \oplus \ldots \oplus M_n \rho^n~.
}
This (reducible) representation corresponds to wrapping $M_{i}$ D2 branes around the positive cycle represented by ${\rho}^{i}$. In the original basis this is
\eqn\thirtysix{
R= \oplus_k N_k R_k \,,
}
where, by brane charge conservation one must have
\eqn\thirtyseven{
N_k = \sum_i M_i n^i_k
}
for the given representation. This then gives rise to a gauge theory with gauge group
\eqn\thirtyeight{
\prod_i U(M_i)~.
}

2) {\it null roots of affine ADE}: these are of the form $N\delta = N(0,1)$ for $N>0$ where ${\delta}=d_{0}e_{0}+{\sum}_{i=1}^{r}d_{i}e_{i}=(0,1)$, and corresponds to the $H_{0}$ class of $ALE$. These correspond to having $N$ $D3$-branes transverse to $ALE$. The dimension of the vector space associated to the $k$-th node of the quiver is $Nd_k$ and the gauge theory has gauge group $G=\prod_k U(Nd_k)$. In contrast with the previous representations, these null branches have nonzero dimension, the space being associated to the motion of $D3$ in the transverse space, as given by the hyper-K\'ahler quotient. The moduli space (Higgs branch) is isomorphic to $N$-fold symmetric products of $ALE$.

3){\it roots of the form $\Delta^+ + N\delta$}: these are $D3-D5$ boundstates. They give rise to a gauge group
\eqn\thnine{
U(N) \times \prod_i U(M_i)~,
}
where again charge conservation gives
\eqn\forty{
N_k = \sum_i M_i n^i_k + Nd_k\,.
}
Moreover, at a generic point of the moduli space of the $U(N)$ theory, this latter group is broken to $U(1)^N$. Also, because these are boundstates, and the $D5$-branes are stuck at the singularity in $ALE$, the Higgs branch for these theories is just a point. Equivalently, the dimension of the quiver representation is zero.

4){\it roots of the form $- \Delta^+ +N\delta$}: similarly to the case above, these are $D3$-${\bar{D5}}$ brane boundstates.

\listrefs
\end{document}